\documentclass[12pt]{article}
\pdfoutput=1

\usepackage{amsmath,amsfonts,amssymb,amsthm,amssymb,bbm,bm}
\usepackage{pdfsync}
\usepackage{cite}
\usepackage{graphicx,import}
\usepackage[latin1]{inputenc}
\usepackage{hyperref}
\usepackage{empheq}
\usepackage[all]{xy}
\usepackage{stmaryrd}
\usepackage{rotating}
\usepackage{color}  
\usepackage{slashed}
\usepackage{cancel}
\usepackage{array, makecell} %
\usepackage{comment}
\usepackage{tikz-cd}
\usepackage{hhline}
\usepackage{empheq}
\usepackage{subcaption}

\DeclareFontFamily{U}{matha}{\hyphenchar\font45}
\DeclareFontShape{U}{matha}{m}{n}{
      <5> <6> <7> <8> <9> <10> gen * matha
      <10.95> matha10 <12> <14.4> <17.28> <20.74> <24.88> matha12
      }{}
\DeclareSymbolFont{matha}{U}{matha}{m}{n}
\DeclareMathSymbol{\oright}       {2}{matha}{"69}

\def\heq{\,\hat{=}\,}

\numberwithin{equation}{section}

\newcommand{\p}{\partial}

\newcommand{\bit}{\begin{itemize}}
\newcommand{\eit}{\end{itemize}}
\newcommand{\bd}{\begin{description}}
\newcommand{\ed}{\end{description}}

\newcommand{\bc}{\begin{center}}
\newcommand{\ec}{\end{center}}
\newcommand{\lbr}{\llbracket}
\newcommand{\rbr}{\rrbracket}

\newcommand{\R}{{\mathbb R}}

\newcommand{\cB}{{\mathcal B}}
\newcommand{\cL}{{\mathcal L}}

\newcommand{\cF}{{\mathcal F}}

\renewcommand{\sl}{{\mathfrak{sl}}}

\newcommand{\be}{\begin{equation}}
\newcommand{\ee}{\end{equation}}
\newcommand{\bea}{\begin{eqnarray}}
\newcommand{\eea}{\end{eqnarray}}
\newcommand{\bs}{\begin{subequations}}
\newcommand{\es}{\end{subequations}}

\newcommand{\la}{\label}

\newcommand{\f}{\frac}

\def\p{\partial}
\def\bC{C}
\def\bD{\bar{D}}
\def\bP{\bar{P}}
\def\bq{\bar{q}}
\def\bU{\bar{U}}

\def\bF{\bar{F}}

\def\para{\text{\tiny{$\parallel$}}}

\def\heq{\,\hat{=}\,}

\newcommand{\dd}{\mathrm{d}}

\def\ip{\lrcorner\,}

\def\g{\gamma}
\def\d{\delta}

\def\th{\theta}

\def\m{\mu}
\def\n{\nu}

\def\s{\sigma}

\def\om{\omega}
\def\G{\Gamma}

\def\Si{\Sigma}

\def\Om{\Omega}
\def\rd{\mathrm{d}}
\def\pa{\partial }

\newcommand{\scri}{\cal I}

\newcommand{\bmw}{BMSW\ }

\begin{document}

\title{\Large{\bf 
Extended corner symmetry, charge bracket and Einstein's equations 
 }}

\author{ Laurent Freidel$^1$\thanks{lfreidel@perimeterinstitute.ca} , 
Roberto Oliveri$^2$\thanks{roliveri@fzu.cz} ,
Daniele Pranzetti$^{1,3}$\thanks{dpranzetti@perimeterinstitute.ca} , 
Simone Speziale$^4$\thanks{simone.speziale@cpt.univ-mrs.fr}
}
\date{\small{\textit{
$^1$Perimeter Institute for Theoretical Physics,\\ 31 Caroline Street North, Waterloo, Ontario, Canada N2L 2Y5\\ \smallskip
$^2$CEICO, Institute of Physics of the Czech Academy of Sciences,\\
Na Slovance 2, 182 21 Praha 8, Czech Republic\\ \smallskip
$^3$ Universit\`a degli Studi di Udine,
via Palladio 8,  I-33100 Udine, Italy
\\ \smallskip
$^4$Aix Marseille Univ., Univ. de Toulon, CNRS, CPT, UMR 7332, \\13288 Marseille, France
}}}

\maketitle

\begin{abstract}
We develop the covariant phase space formalism allowing for non-vanishing flux, anomalies, and field dependence in the vector field generators.
We construct a charge bracket that
generalizes the one introduced by Barnich and Troessaert and includes contributions from the Lagrangian and its anomaly. 
This bracket is uniquely determined by the choice of Lagrangian representative of the theory. 
We then extend the notion of corner symmetry algebra to include the surface translation symmetries and prove that the charge bracket provides a canonical representation of the extended corner symmetry algebra. 
 This representation property is shown to be equivalent to the projection of the gravitational equations of motion on the corner, providing us with an encoding of the bulk dynamics in a locally holographic manner.
 
\end{abstract}

\newpage
\tableofcontents

\section{Introduction }\la{sec:Intro}

 The notion of corner symmetry algebra represents an essential tool to study the subdivision of space into local subsystems, and their gluing and coarse-graining properties \cite{DonnellyFreidel}. 
 It also permits the reformulation of geometrical observables in an algebraic language amenable to quantization.  
 Overall, the program of \emph{local holography} provides a clear pathway toward quantization, where the geometry is encoded into a charge algebra, the quantum kinematics into the algebra's representations, and the quantum dynamics into the algebra's fusion properties. 
 So far, the study of the corner symmetry algebra has been limited to the kinematical sector, namely to transformations that do not move the corner \cite{Freidel:2015gpa, Freidel:2016bxd, Freidel:2018pvm, Freidel:2019ees, Freidel:2019ofr, Freidel:2020xyx, Freidel:2020svx, Freidel:2020ayo, Donnelly:2020xgu}.  
 The natural next step in this program
 is to consider the so-called  extended corner symmetry group
 which also includes the possibility to translate the corner along its normal directions.
 
 The previous investigations are relevant to constructing a  Hilbert space of quantum geometry labeled by quantum numbers associated with the corner symmetry algebra representation. 
  From the perspective of local holography, the gravitational dynamic is encoded in general conservation laws, called flux-balance laws, for the corner charges.  While the exact form of the extended corner symmetry group can depend on the gravity formulation (see \cite{Freidel:2020xyx, Freidel:2020svx} for a classification), this group always contains a subfactor that descends from the whole diffeomorphism group. This subfactor is the extended symmetry group of the Einstein-Hilbert formulation of gravity.  Its corner component, studied in  \cite{DonnellyFreidel, Donnelly:2020xgu}, 
is given by the semi-direct product of diffeomorphisms tangent to the sphere and surface boosts. The fully extended version also contains sphere's normal translations and it is given by the semi-direct sum\footnote{\la{1}The corner subalgebra is $ \mathfrak{g}_S =  \mathrm{diff}(S)\oright \mathfrak{sl(2,\mathbb{R})}^S$ where $G^S$ denotes the sets of functions from $S\to G$. In general the full symmetry group contains a $\mathfrak{gl(2,\mathbb{R})}^S$ factor. 
Only $\sl(2,\R)^S$ is represented non-trivially in Einstein-Hilbert, while the diagonal component is pure gauge.   }
\be\label{gs}
\mathfrak{g}^{\rm ext}_S = \left( \mathrm{diff}(S)\oright \mathfrak{sl(2,\mathbb{R})}^S\right) \oright (\R^2)^S \,.
\ee
This symmetry algebra has been shown, in a very recent work of Ciambelli and Leigh \cite{Ciambelli:2021vnn}, to be the maximal closed subalgebra of the full bulk diffeomorphism group associated to isolated corners. From our perspective, this means that its representation theory should appear as a  \emph{universal} component of any quantization of gravity.

We are interested in 
a canonical representation of this algebra using covariant phase space methods \cite{Kijowski1976ACS,  Crnkovic:1986ex,Ashtekar:1990gc,Lee:1990nz,Barnich:1991tc,Wald:1999wa}. This 
introduces a non-trivial problem because the normal translations move the surface, and the corresponding infinitesimal Hamiltonian charges are in general non-integrable.
Suppose one restricts the normal translations to be tangent to a timelike or null boundary. In that case, the non-integrability of charges is related to a non-vanishing symplectic flux along the boundary and the corner symmetry algebra reduces to a  boundary symmetry algebra.   
Understanding the meaning of non-integrable charges in the presence of boundaries is not a new problem and has been the subject of extended studies, starting with Ashtekar et al. \cite{Ashtekar:1981bq,Dray:1984rfa,Ashtekar:1990gc},  
Wald--Zoupas \cite{Wald:1999wa}, Barnich et al. \cite{Barnich:2001jy,Barnich:2004uw,Barnich:2007bf, Barnich:2009se,Barnich:2010eb, Barnich:2011mi, Barnich:2013axa}, Compere et al. \cite{Compere:2018ylh, Compere:2020lrt} in the context of asymptotic infinity, by
Donnay et al. \cite{Donnay:2015abr,Donnay:2016ejv,Donnay:2019jiz},
 Hopfm\"uller and one of us  \cite{Hopfmuller:2016scf,Hopfmuller:2018fni}, Grumiller et al. \cite{Adami:2020amw, Grumiller:2019ygj, Grumiller:2020vvv},  
Chandrasekaran et al. \cite{Chandrasekaran:2018aop, Chandrasekaran:2020wwn}, from the perspective of finite boundaries.
Recent developments include
the presence of a cosmological constant \cite{Compere:2019bua, Compere:2020lrt, Alessio:2020ioh, Fiorucci:2020xto}, the issue of field dependency in  lower-dimensional gravity by Adami  et al.
\cite{Adami:2020ugu} and Ruzziconi and Zwickel \cite{Ruzziconi:2020wrb}, and the relationship between flux and edge mode dynamics along null surfaces by Wieland \cite{Wieland:2020gno,Wieland:2021eth}.
Our paper generalizes these analyses to the case where the normal translations span the full two-dimensional subspace of normal directions, not just a one-dimensional subspace along a given boundary. The results obtained here were already  announced in \cite{Freidel:2021yqe}.

The presence of non-vanishing symplectic and more generally Hamiltonian fluxes, leads to two issues that one has to resolve: one technical and one conceptual. 
The technical issue stems from the fact that when Hamiltonian fluxes are present, it is not 
clear how to extract from the formalism a Hamiltonian charge. 
More precisely, there is an ambiguity on defining a split between an integrable term and a Hamiltonian flux term of the field space one-form  obtained by
contracting an infinitesimal diffeomorphism along the symplectic two-form.
The conceptual issue is that in the presence of fluxes, the physical system under consideration is an open system that does not possess a canonical Poisson bracket.

In the ``Belgian school'' approach \cite{Barnich:2011mi, Barnich:2010eb,Barnich:2013axa, Compere:2018ylh, Compere:2020lrt}  the choice of split between the Hamiltonian charge and the flux component is left arbitrary.
Barnich and Troessaert have proposed a resolution of the conceptual issue though. They introduced, initially in the context of asymptotically flat spacetimes, a flux-dependent bracket \cite{Barnich:2011mi} which reduces to the usual Poisson bracket when there are no fluxes.
This bracket leads to two concerns: first, the charge algebra defined by this bracket admits field-dependent cocycles, which depend on the split ambiguity. And second, the Jacobi identity for the Barnich--Troessaert  bracket \cite{Barnich:2011mi} has to be postulated rather than proven.

We are proposing a new perspective on these questions which builds on the Wald--Zoupas approach \cite{Wald:1999wa} and on the  recent work of Chandrasekaran and Speranza \cite{Chandrasekaran:2020wwn}.
It also builds upon the results of \cite{Freidel:2020xyx} which proved that one can assign a unique symplectic potential given a specific  Lagrangian.
Our first main result is the construction, given a Lagrangian, of an unambiguous split between charge and flux, with the Hamiltonian flux  chosen to be Noetherian,  and a definition of a bracket that allows for the inclusion of all normal translations and which satisfies the Jacobi identity from first principles.

To describe our result, let us introduce the notion of Lagrangian equivalence class $[L]=[L+\rd \ell]$ to be the equivalence class of Lagrangians modulo boundary terms.   Our approach relies on the fact \cite{Freidel:2020xyx, anderson1992introduction, Lee:1990nz} that, given a representative Lagrangian $L$ in a given class, we can uniquely construct  its symplectic potential $\theta^L$,
the Noether charge $Q_\xi^L$ and the Noetherian flux $\cF_\xi^L$  associated with a vector field $\xi$.
The choice of Noether charge and Noetherian flux resolves the split ambiguity described earlier.
The centerpiece of our construction is a definition of a bracket $\{\cdot,\cdot\}_L$ uniquely associated to  $L$ and free of any boundary conditions and corner ambiguities.
This bracket provides a generalization of the Barnich--Troessaert bracket
 \cite{Barnich:2011mi}, in that it is also defined for the extended corner symmetry group at finite corners, and it takes into account the presence of Lagrangian anomalies (as defined in Section \ref{sec:Nth}).

 The second main result of the paper is to show that, when evaluated on-shell, the new charge bracket  provides a 
 representation of the extended corner symmetry algebra.
 Moreover the logic can be inverted and we show that demanding  this bracket to form a faithful representation of extended corner symmetry algebra 
 implies the validity of the Einstein's equations, as explicitly shown for null infinity in \cite{Freidel:2021yqe}. This statement is encoded in our main formula
 derived in Section \ref{sec:constraints}
 \be 
  \{ Q^L_\xi, Q^L_\chi\}_L+Q^L_{\lbr\xi,\chi\rbr}+ \int_S\iota_\xi  C_\chi =0\,,
 \ee 
 where $C_\xi$ is the Einstein constraint 3-form along $\xi$ (see \eqref{Ccon}). 
 This
 reveals the locally holographic nature of gravity at any given corner, by recasting Einsteins's equation into an algebraic statement.

The fact that we have a bracket that depends on the choice of Lagrangian sounds surprising at first.
However, it is well known that when we have a Lagrangian system in the presence of a boundary one needs to specify boundary conditions to make the system closed and thus obtain a canonical bracket. This specification of boundary conditions is a crucial ingredient in the definition of the non-leaking phase space. The integrable charges and the canonical bracket, therefore, depend on the choice of boundary condition.

An important point, realized already by Harlow and Wu in \cite{Harlow:2019yfa},
is that the specification of boundary conditions corresponds to a choice of boundary  Lagrangian. In other words,  a selection of Lagrangian representative $L$ within a given class specifies a boundary condition.
The way this work is as follows: 
a choice of boundary conditions amounts to a choice of which phase space variables $q^i$, forming a Lagrangian submanifold, to keep fixed at the spatial boundary $\Gamma$. In a Lagrangian framework, this requires to identify also their conjugate momentum $p_i$, so that the symplectic potential $\theta=p_i \d q^i$ vanishes on $\Gamma$ once the boundary condition 
\be\la{pqG}
\d q^i \stackrel{\Gamma}{=}0
\ee 
is imposed. Changing the boundary conditions is done by a canonical transformation implemented at the Lagrangian level by adding a boundary Lagrangian.
 In this way, the Poisson bracket for a given choice of boundary conditions is uniquely determined by the choice of boundary Lagrangian, while the symmetries of the system appear as canonical transformations that do not change the boundary conditions. In our language, a Lagrangian $L$ determines the symplectic potential $\theta^L$ and the boundary condition 
 ${\cal B}_L: \,\theta^L=0$. The canonical bracket therefore depends on the choice of Lagrangian. 
 An important technical aspect of our analysis, which was missing in  \cite{Harlow:2019yfa}, is that we have extended the construction of Noether charge and flux for Lagrangians admitting anomalies (see also \cite{Chandrasekaran:2020wwn} and~\cite{Margalef-Bentabol:2020teu})
 and to the case of leaking phase spaces without restricting to Dirichlet boundary conditions as  in  \cite{Harlow:2019yfa}.
 In Section \ref{sec:NvH} we prove that the Lagrangian bracket we construct reduces to the canonical bracket when ${\cal B}_L$ is imposed. We also show that  the Noether charge $Q^L_\xi$ reduces to the Hamiltonian charge.\footnote{In the following, we drop the index $L$ denoting the dependence of all the phase space quantities on the choice of Lagrangian in order to lighten the notation. It will later reappear in the definition of the bracket in Section \ref{sec:IBra} and in Section \ref{sec:NvH}.}

Finally, in Section \ref{sec:4} we study the  momentum map of the  Einstein--Hilbert formulation of gravity
representing the extended corner algebra \eqref{gs}, and we compare our results with \cite{Ciambelli:2021vnn}.
To investigate the physical meaning of the charges associated with this quasi-local symmetry, we study their limit to future null infinity in Bondi coordinates. This reveals that the extended corner symmetry algebra tends to the BMSW group, the generalization of the BMS group we recently proposed in \cite{Freidel:2021yqe}. In particular, the corner symmetry charges tend to the
arbitrary tangent diffeomorphism  and the local Weyl rescaling  part of the \bmw Lie algebra, the new normal charges tend to the super-translation charges, and one of these quasi-local charges vanishes in the limit.

{The paper starts with the introduction of the concept of anomaly in the covariant phase space formalism in Sec.~\ref{sec:2}. Section \ref{sec:3} is devoted to the construction of the new charge bracket, and its relation to the flux-balance laws and Einstein's equations. Section \ref{sec:4} extends the kinematical corner symmetry algebra to include the dynamical contribution, and discuss its application at null infinity. We conclude with Sec.~\ref{sec:concl}. The technical steps to prove the results in Sec.~\ref{sec:2} and Sec.~\ref{sec:3} are collected in the appendix.}

\section{Covariant phase space in the presence of anomalies}\label{sec:2}

The natural arena to develop the covariant phase space formalism is the jet bundle, where fields {and their derivatives} can be viewed as section of a fiber bundle over the {base} 
manifold provided by the spacetime $M$.  We introduce a set of local coordinates $(x^\m, \varphi^i)$ on $U_M\times  U_F$, where $U_M, U_F$ are open sets on $M$ and the fiber $F$, respectively. By taking a section of the fiber, we can view fields as maps  ${ \varphi}: U_M \to U_F $ given by $ x\to\varphi^i(x)$.  On the jet bundle then, the horizontal derivative provides a  notion of spacetime differential, which we denote by $\rd$, and the vertical derivative a notion of field space differential, which we denote by $\delta$. These are  two key ingredients of the bi-covariant Cartan's calculus that we are going to review briefly in Section \ref{sec:Nth}. More details can be found in \cite{anderson1992introduction, Compere:2018aar, Freidel:2020xyx}.
The next crucial element is a Lagrangian 
{top} form $L$, which defines the physics of the fields and the symmetries of spacetime,
{whose deep relation is the subject of our investigation.}

\subsection{Symplectic Flux}

As shown in \cite{Lee:1990nz, Freidel:2020xyx}, using  Anderson's homotopy operators \cite{anderson1992introduction,Compere:2018aar}, it is possible to associate a unique symplectic potential $\theta =\theta^{L}$ to a given Lagrangian $L$.
This potential is such that 
\be \label{fundamental}
\delta L = \rd \th - E,
\ee where $\delta$ is the variational Cartan differential and $E$ are the equations of motion.\footnote{For metric gravity, we have $E=  G^{\m\n}\delta g_{\m\n}\sqrt{|g|}\rd^4x$, where $G^{\m\n}$ is the Einstein tensor.}
As emphasized in \cite{Freidel:2020xyx}, this identity should \emph{not} be taken as a definition of the symplectic potential, but as a consequence of the definition of $\theta$ from $L$.
The (pre)-symplectic form\footnote{It is symplectic before we impose equations of motion and pre-symplectic upon doing so.} of the theory is a variational 2-form, obtained after integration of $\delta \theta$ on a codimension-$1$ hypersurface $\Si$,
 \be\label{sympf}
 \Omega :=\int_\Si \delta \theta. 
\ee
In this paper, we assume that $\Sigma$ is a spacelike or null  hypersurface with  boundary $S :=\pa \Sigma$.
The boundary is not required to be connected, 
and can generally be decomposed as a disjoint union 
$S=\cup_i S_i$, where the individual codimension-2 components $S_i$ can be surfaces of arbitrary genus, see Fig.~\ref{boundaries}.
Most of the results presented below apply to any topology of $\Sigma$ and $S$.
\begin{figure}[htb]
\centering
\begin{subfigure}[htb]{60mm}
\centering
\includegraphics[height=39mm]{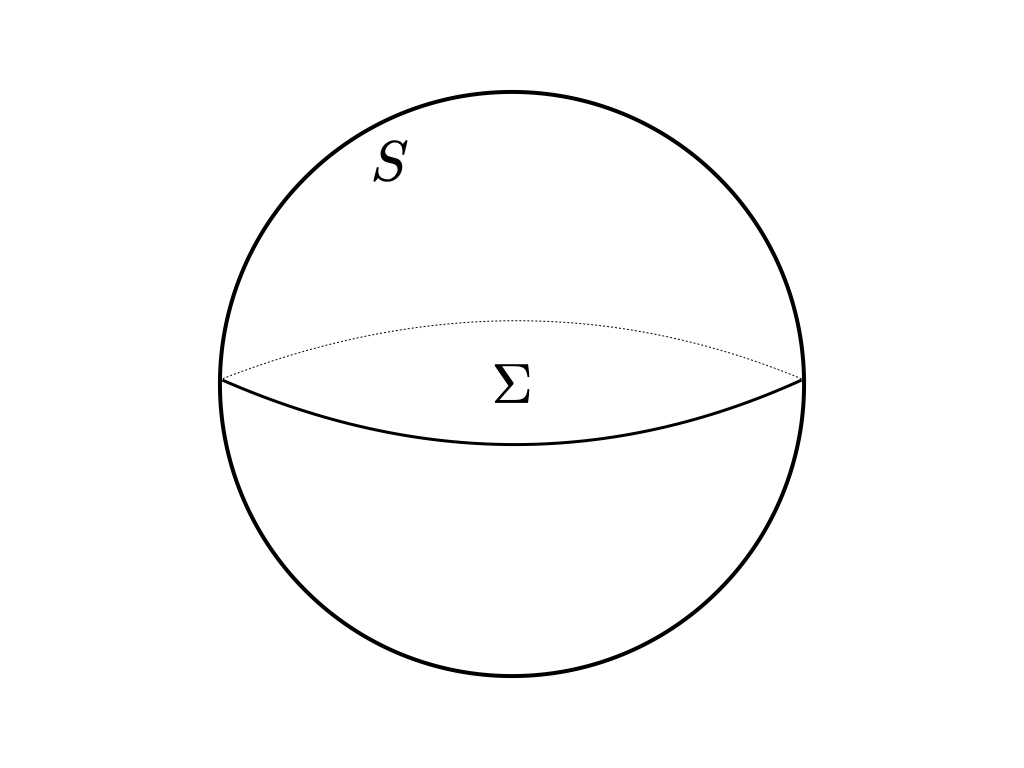}
\end{subfigure}
\hspace*{14.mm}
\begin{subfigure}[htb]{60mm}
\centering
\includegraphics[height=39mm]{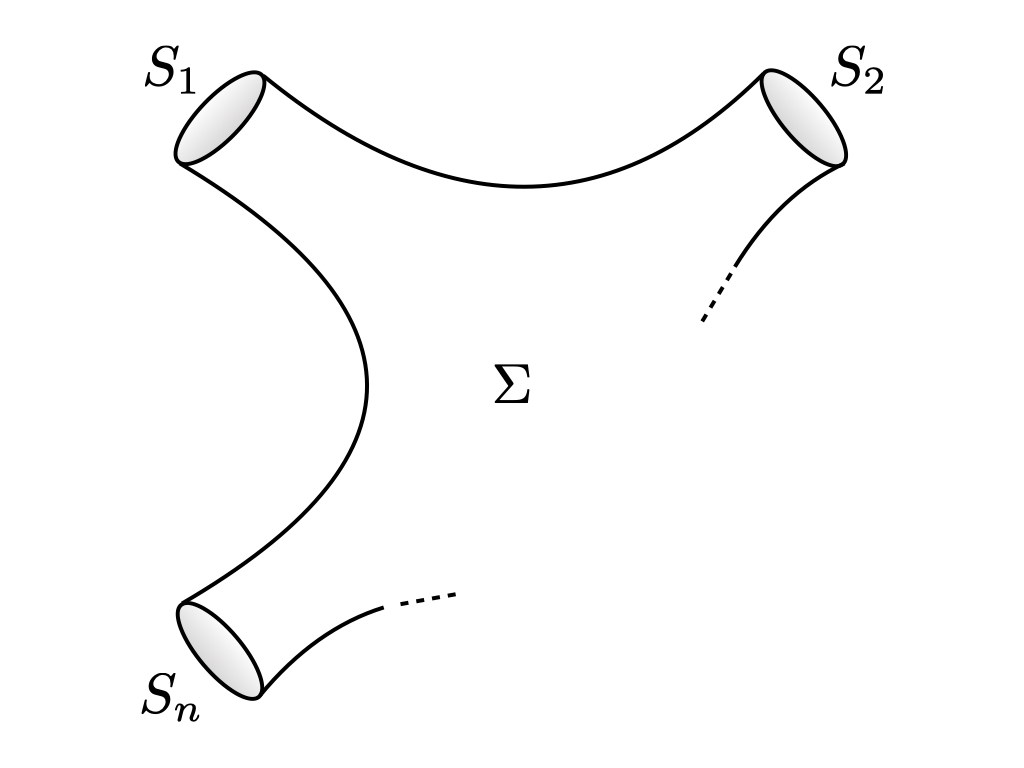}
\end{subfigure}
\caption{\small{\emph{Different topologies of the hypersurface  $\Sigma$.}}}
\la{boundaries}
\end{figure}
 
One exception is the construction of the bracket in Section \ref{sec:IBra}, for which we will restrict the topology of $\Sigma$ to insure that closed 3-forms are exact. The simplest set-up is when $\Sigma$ is a 3-ball and $S$ is a {single} two-sphere.
We refer to this sphere as the entangling sphere, since this is usually the place where the entanglement entropy of the two regions separated by $S$ lives. 
If a black hole is present, the region $\Sigma$ connected to spacetime infinity 
cannot be a 3-ball, but additional internal boundaries are needed to exclude  singularities.

As a consequence of the fundamental relation \eqref{fundamental}, 
the symplectic current $\omega := \delta \theta$ is conserved on-shell since $\rd \omega =\delta E$.
However, this does \emph{not} necessarily mean that the symplectic form \eqref{sympf} is 
{independent of $\Si$}. The conservation of the symplectic form depends on what happens at its boundary.
The simplest case is when the 
{topology is such that all $\Si$'s are hinging at the same codimension-2 corner; see left panel of Fig~\ref{Foliations}. 
In the following we are interested in a more general situation in which there is a non-trivial time development of the slices $\Sigma$ that translates the boundary $S$.  
When this happens, the initial and final slices are connected by a 3d boundary $\Gamma$, which represents the time development of the corner $S$; see right panel of Fig~\ref{Foliations}. 
In this case, \eqref{sympf} can differ at the initial and final slices due to the presence of symplectic flux going through the 3d boundary.

To be concrete, let us assume that we choose a collection of time-dependent 
slices $\Sigma_t$ with boundaries $S_t$ (they do not have to be complete Cauchy hypersurfaces). If we denote $\xi =\pa_t$ the time translation vector field
and $\Phi_{\xi} =\exp \xi$ the associated diffeomorphism, we have that $\Sigma_t = \Phi_{t\xi}(\Sigma)$ and similarly $S_t = \Phi_{t\xi}(S)$, while $
\pa_t \Phi_{t\xi} = \xi \circ \Phi_{t\xi}
$, {and $\Si$ with boundary $S$ are reference initial slices}.
The time dependent symplectic form is 
\be \label{actpassive}
\Omega_t:= \int_{\Sigma_t}\!\!\! \omega 
= \int_{\Sigma} \Phi_{t\xi}^*(\omega)\,,
\ee
where $\Phi^*_{t\xi}$ denotes the pull back of forms by the diffeomorphism $\Phi_{t\xi}$. This formula shows that we can view the active displacement of a hypersurface inside spacetime in a passive way as the transformation of fields leaving on a single slice.

The two cases evoked above can now be represented as follows:
either the  time foliation leaves the boundary fixed, i.e. $S_t=S$,  
or the diffeomorphism $\Phi_{t\xi}$ moves the boundary surface and the union $\cup_t S_t$ forms a foliation of the codimension-$1$ {boundary} $\Gamma$ (see Fig. \ref{Foliations}).
In the first case, we have that the symplectic structure is conserved in time and that the action of $\xi$ admits a canonical  representation on the gravity phase space. 
This is the case studied in \cite{DonnellyFreidel,Donnelly:2020xgu}, where the sets of admissible vector fields form a corner symmetry algebra with the semidirect sum structure $\mathfrak{g}_S = \mathrm{diff}(S)\oright \mathfrak{sl(2,\mathbb{R})}^S$.
In the second case, the symplectic form is not conserved {in general}. Its non-conservation is encoded into 
the presence of a non-zero  flux at the boundary $S$
\be\label{sympflux}
\pa_t \Omega_t \heq \delta \cF^{\theta}_{\xi}- \cF^{\theta}_{\delta \xi}, 
\ee
where $\cF^{\theta}_\xi := \int_S \iota_{\xi} \theta  $ will be referred to as the \emph{symplectic flux} associated with $\xi$,
and we use the symbol 
$\heq$ when   the equations of motions are imposed, namely  $E\heq 0$. 
Eq.~\eqref{sympflux}  is the infinitesimal version of Stokes' theorem:
\be\label{Stokes}
\Om_{t}-\Om_{0}\heq \Om_{\G_t }\,,
\ee
where  $\Gamma_t$ is  the portion of $\Gamma$ between $\Sigma=\Sigma_0$ and $\Sigma_t$, and
$\Om_{\G_t } =\int_{\G_t} \omega$ is the integrated symplectic flux which measures the loss of symplectic information going through the boundary.
\begin{figure}[htb]
\centering
\begin{subfigure}[htb]{74mm}
\centering
\includegraphics[height=39mm]{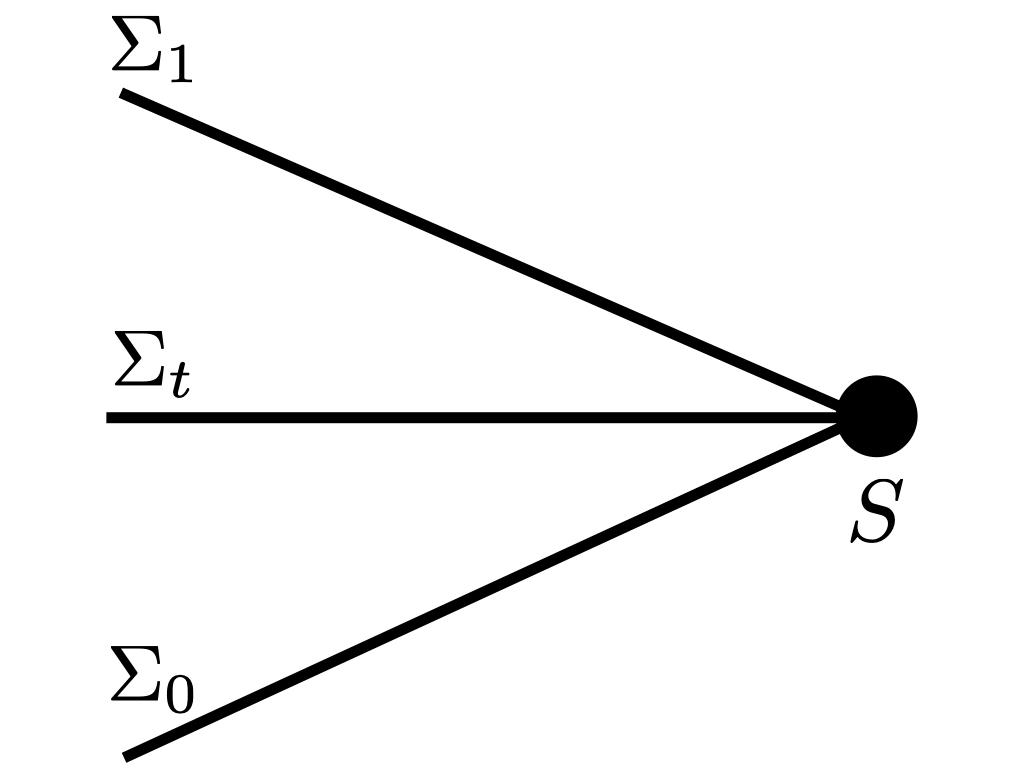}
\caption{Entangling sphere $S=S_t$ where all the  spacelike Cauchy hypersurfaces $\Sigma_t$ are joined.}\la{Wedge}
\end{subfigure}
\hspace*{14.mm}
\begin{subfigure}[htb]{74mm}
\centering
\includegraphics[height=39mm]{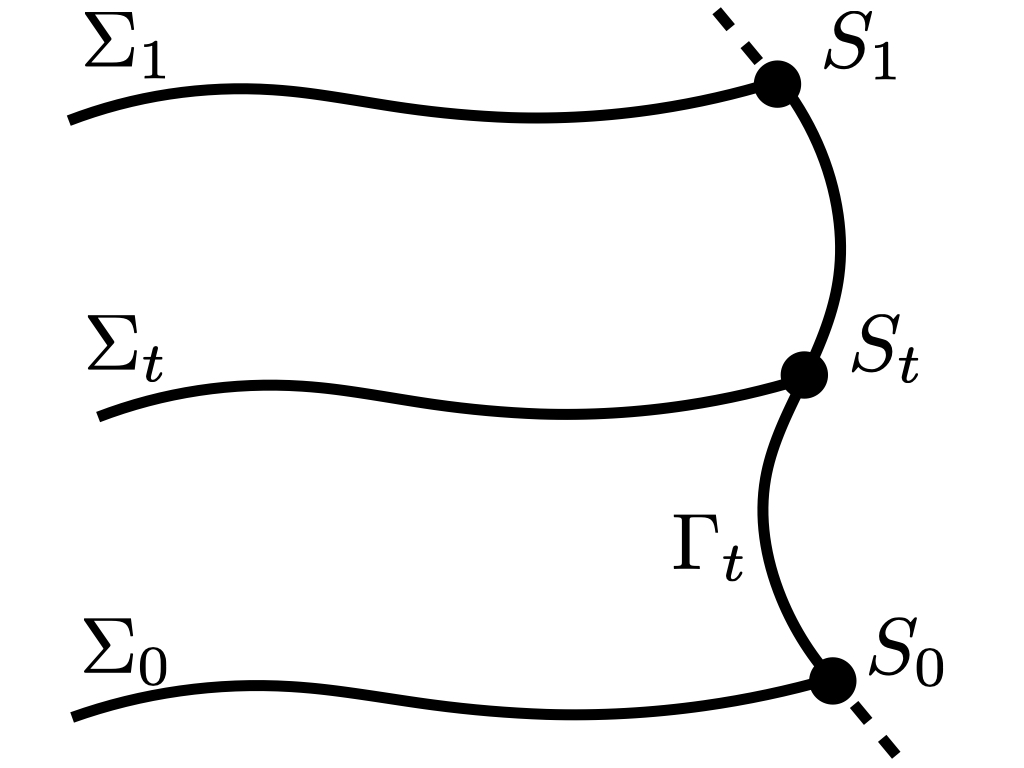}
\caption{Codimension-1  timelike or null boundary $\Gamma$ foliated by spacelike Cauchy hypersurfaces   $\Sigma_t$ through the union $\cup_t S_t$.}\la{Gamma}
\end{subfigure}
\caption{\small{\emph{Foliation leaves $\Sigma_t$ with boundaries $S_t$.}}}
\la{Foliations}
\end{figure}
To prove \eqref{sympflux}  one uses \eqref{actpassive} and the fundamental variational formula \eqref{fundamental} to establish that 
\bea
\pa_t \Omega_t &=& \int_\Sigma \cL_\xi \omega
= 
\int_\Sigma \cL_\xi \delta \theta 
=  \int_S  \iota_\xi \delta\theta
+ \int_\Sigma \iota_\xi  \delta \rd \theta \cr
&=& \delta \left( \int_S \iota_\xi \theta \right)
-\int_S \iota_{\delta \xi}  \theta
+ \int_\Sigma \iota_\xi {\d }E.
\eea

\subsection{Boundary Lagrangian shift}

As recalled in the introduction, the nature of the symmetry algebra has been the subject of extensive studies by 
\cite{Ashtekar:1981bq,Dray:1984rfa,Ashtekar:1990gc, Wald:1999wa,Barnich:2001jy,Barnich:2007bf, Barnich:2009se,Barnich:2010eb, Barnich:2011mi, Barnich:2013axa,Compere:2018ylh, Compere:2020lrt,Alessio:2020ioh, Fiorucci:2020xto} in the context of asymptotic infinity, and by \cite{Adami:2020amw, Grumiller:2019ygj, Grumiller:2020vvv,Donnay:2015abr,Donnay:2016ejv,Donnay:2019jiz, Hopfmuller:2016scf,Hopfmuller:2018fni, Chandrasekaran:2018aop} from the perspective of finite null boundaries. Since the understanding the nature of the corner symmetry algebra in the presence of non-zero flux is the goal of this paper, we extend previous analyses to the case where the the normal translations are not restricted to be along a given boundary.

The first key ingredient to achieve this is the fact that the symplectic potential is uniquely determined by the Lagrangian \cite{Freidel:2020xyx}, and the following important consequence.
Let us suppose that we modify the original Lagrangian by a boundary term, so that 
the two Lagrangians have the same equations of motion. We then get two distinct symplectic potentials $\theta$ and $\theta'$, respectively associated to $L$ and $L'=L+\rd \ell$.
The difference between the two symplectic potentials can be understood as the equations of motion for the boundary Lagrangian.
In other words, we have that \cite{Freidel:2020xyx}
\be\label{L'}
L'-L=\rd \ell,\qquad \theta' -\theta = \delta \ell - \rd \vartheta ,
\ee
where $\ell$ is the boundary Lagrangian, while  $\vartheta$ is the corner symplectic potential 
for  $\ell$. 

The fact that one can assign a unique symplectic form to the pair $(L,\ell)$ is an important point which was overlooked in the early references, such as \cite{Wald:1999wa}, where this type of changes in Lagrangian and symplectic potential are often considered to be ambiguities of the covariant phase space.  Instead, in \cite{Freidel:2020xyx} it was shown to be a feature of a gauge theory that 
can lead to different representations of the corner symmetry algebra.  More precisely, 
the new insight of \cite{Freidel:2020xyx} is that the symplectic form is modified by the addition  of a boundary Lagrangian through 
its corner symplectic potential
\be\label{shiftO}
\Omega'= \Omega - \int_{S} \delta \vartheta. 
\ee

A clear way to appreciate the relevance of the property \eqref{shiftO} is to realize its  key role in the asymptotic renormalisation of  the  charges
\cite{Papadimitriou:2005ii,Compere:2008us, Compere:2018ylh, Freidel:2019ohg,Compere:2020lrt, Freidel:2021yqe}, as it  allows one to reabsorb the divergences of the covariant symplectic form entirely in terms of the corner symplectic potential.
Augmenting the definition of the covariant phase space with this prescription, one gets a better handle of physical quantities than working with an equivalence class of symplectic potentials, as clearly exhibited in \cite{Freidel:2021yqe}.
We remark also that the transformations \eqref{L'} were originally shown in \cite{Harlow:2019yfa}  to be related to a choice of boundary conditions;  we will come back to this important point more in detail in Section \ref{sec:NvH}.

\subsection{Noether theorem, Noether charges and  Anomalies}\la{sec:Nth}

Once the pair $(L,\theta)$ is chosen, we can apply the Noether theorems  to construct the symmetry charges $Q_\xi$.  By the second Noether theorem, the Noether charges for local gauge symmetries  are really 
corner charges  on-shell. This means that $Q_\xi =\int_S q_\xi$, where $q_\xi$ is a codimension-$2$ form called the \emph{charge aspect}.
It is important to appreciate that the charge aspect and the charge $Q_\xi$ are \emph{uniquely} determined from the choice of a Lagrangian.\footnote{ With the use of the homotopy operators \cite{Freidel:2020xyx}.}
This is what we call the Noether charge associated with $(L,\theta)$. Of course, an important part of the construction requires understanding how the charges and fluxes transform under changes of Lagrangians giving the same equations of motion.

The explicit construction  requires
 introducing the notion of field space interior product and Lie derivative.
Given a vector field $\xi$ in spacetime, we denote ${\cal L}_\xi$ the Lie derivative and 
 $\delta_\xi$ the corresponding field variation.\footnote{ Strictly speaking the proper notation should be  $ I_{{\cal L}_\xi } = {\cal L}_\xi \ip $ which refers to the interior product along the field variation. For simplicity we use $I_{\xi}:=I_{{\cal L}_\xi }$ and $\delta_\xi :=\delta_{{\cal L}_\xi }$.}
 We also denote $\iota_\xi$ the vector field contraction on spacetime forms 
and $I_\xi$ the field space contraction on field space forms.
The field space contraction $I_\xi$ is such that when acting on simple forms $\delta \phi$,
with $\phi$ a scalar,
it gives the variation $I_\xi \delta \phi = \delta_\xi \ip \delta \phi = \delta_\xi \phi$.

In the following we will be interested in field-dependent diffeomorphisms, hence $\d\xi\neq 0$.
In this case, it is necessary to consider also the operator $I_{\delta \xi}$ which denotes 
 the field contraction along a  form-valued vector. While $I_\xi$ is a derivation that lowers the form degree by $1$, $I_{\delta \xi}$ is a derivation of degree $0$, like $\delta_\xi$. And while 
 the contractions commute\footnote{All commutators are bi-graded commutators (see Appendix A in \cite{Freidel:2020xyx} for more details).} $ [I_{\chi},I_{\xi}]=0$, the form-valued contraction satisfies $[I_{\chi}, I_{\delta \xi}]= I_{\delta_\chi \xi}$; see \cite{DonnellyFreidel} and \cite{Speranza:2017gxd} for  more details on these notations.
The Lie derivatives and interior products are related by Cartan's 
formula and its field space analog  
\be
{\cL_\xi}= \rd \iota_\xi +\iota_\xi \rd, \qquad \delta_\xi = \delta I_\xi +I_\xi \delta.
\ee
For the gravitational field space, we have $\delta_\xi g_{\m\n} = \cL_\xi g_{\m\n}$.

In this work, we are specifically interested in the construction of charges associated with Lagrangians which are {\it semi-covariant}
under diffeomorphisms. 
Semi-covariant under diffeomorphisms simply means that the variation of the Lagrangian  is a total derivative,
\be\label{deltal}
\delta_\xi L =\rd \ell_\xi\,.
\ee
We also  call \emph{covariant} the  Lagrangians for which 
$\ell_\xi =\iota_\xi L$. This is the case when the Lagrangian transforms under field diffeomorphism as a top form:
$\delta_\xi L=\cL_\xi L= \rd (\iota_\xi L).$

Noether's remarkable\footnote{It is quite remarkable that Noether's paper is already about the covariant phase space and it already contains all the elements necessary to construct the bicovariant calculus!
It took more than 60 years for the rest of theoretical physics to catch up to the depth of this paper. It is only in the very recent years that we are starting to outgrow it. For the wonderful history see \cite{Kosmann,kosmannschwarzbach2020noether}.} paper \cite{Noether:1918zz} establishes two fundamental results from this covariance.
First, it shows that the equations of motion have to satisfy a Bianchi identity associated with each gauge symmetry and this implies, in our language, that there exists a constraint $C_\xi$, which vanishes when $E \heq 0$, given by 
\be
I_\xi E =\rd C_\xi. 
\ee 
Furthermore,  it also establishes\footnote{To continue on the historical note. The formula for the Noether charge that includes the boundary term $\ell_\xi$ was first proposed by Bessel--Hassel in 1921 
\cite{Bessel-H,Bessel-HT}.} that the Noether current is the sum of  the constraints and a total differential 
\be\label{Noether}
j_\xi:= I_\xi \theta - \ell_\xi = C_\xi +\rd q_\xi. 
\ee

The main limitation of the usual application of  Noether's theorems is the assumption that 
the action of a diffeomorphism in field space coincides with the Lie derivative in spacetime, namely that  $\d_\xi=\cL_\xi$. In this case, we say that the formalism is free of anomalies.
In the presence of gauge-fixings and boundary conditions, this is not necessarily true anymore.
For instance, the use of the Gibbons--Hawking boundary term in the definition of the Lagrangian, to enforce Dirichlet's boundary conditions, introduces Lagrangian anomalies {because of the appearance of the boundary normal}. Moreover,
anomalies naturally appear in the presence of background structures that break covariance.
 Such structures are necessary to define the notion of infinity.
 For instance, suppose that we choose a foliation of our spacetime such that the normal to the foliation is 
 the one form $N_\mu\rd x^\mu=\rd r$. The coordinate $r$ is a preferred coordinate (which allows, for instance, the definition of  the asymptotic limit $r\to \infty$ {or of the factor for the the conformal compactification}) that can be used to define a 
vector field $N^\mu = g^{r\mu}$ normal to the foliation.
The transformation of this vector under diffeomorphisms is  given 
 by $\cL_\xi N^\mu = [\xi,N]^\mu$. The field transformation remembers, on the other hand, that this vector is the component of a tensor.\footnote{Explicitly, we have $\d_\xi N^\m= \cL_\xi g^{r\m}= \xi^\n \p_\n g^{r\m}-g^{r\n}\p_\n \xi^\m - g^{\n\m}\p_\n \xi^r=  [\xi,N]^\m - N^\mu \p_r \xi^r$.}  That creates an anomaly given by
 $\Delta_\xi N^\mu=(\d_\xi-\cL_\xi) N^\mu =- N^\mu \p_r \xi^r$.
 
 This means that the derivation of the Noether charges, Noetherian fluxes and Poisson brackets needs to be revisited in the presence of anomalies.
This is especially crucial in order to define symplectic renormalization \cite{Freidel:2021yqe}.
The goal of this section is to further develop the consequences of anomalies in the covariant phase space formalism.

 The concept of anomaly in the covariant phase space formalism was 
studied by one of us in \cite{Hopfmuller:2018fni}. {Earlier works on anomalies in covariant phase space formalism include \cite{Tachikawa:2006sz,Azeyanagi:2015gqa}.}
The fact that boundary Lagrangians introduce anomalies 
was also noted by Harlow and Wu in \cite{Harlow:2019yfa}.
Moreover, the symplectic formalism in the presence of anomalies has been recently 
developed by  Chandrasekaran and Speranza  in \cite{Chandrasekaran:2020wwn}. In this work we  develop further their results and generalize their analysis.

 Given a form in spacetime and field space $\omega $, we define its \emph{anomaly} by the difference between the field space action and the spacetime Lie derivative:
\be\la{ano}
\Delta_\xi \omega, \quad\mathrm{with}\quad
\Delta_\xi:=(\delta_\xi -{\cal L}_\xi - I_{\delta \xi}).
\ee  
The anomaly operator $\Delta_\xi $ is a graded derivation of degree $0$  which satisfies the commutation relations
\be\label{Com1}
\rd \Delta_\xi =\Delta_\xi \rd\,\qquad
\delta \Delta_\xi = \Delta_\xi \delta + \Delta_{\delta \xi}\,.
\ee
When acting on the Lagrangian, a scalar in field space, the anomaly is simply $\Delta_\xi L = \delta_\xi L  -{\cal L}_\xi L$.

As said above,  following Noether and Bessel-Hassel \cite{Noether:1918zz,Bessel-H,Bessel-HT}, we are interested in theories with a semi-covariant Lagrangian, i.~e. such that the Lagrangian anomaly is a pure boundary term. 
It then follows that the symplectic anomaly is also determined by the sum of a pure boundary term and a pure variational term: 
\be\label{covarianceprop}
\Delta_\xi L= \rd a_\xi,
\qquad
\Delta_\xi \theta = \delta a_\xi -a_{\delta \xi} + \rd A_\xi\,.
\ee
The first equation is the definition of semi-covariance, the second is a postulate for the most general anomaly allowed by the relation $\delta L =\rd \theta -E$.\footnote{
One simply evaluates
\bea
\Delta_\xi( \delta L -\rd \theta) =
\delta \Delta_\xi L -\Delta_{\delta \xi}L - \rd \Delta_\xi \theta= \rd(\delta  a_\xi -  a_{\delta \xi} -\Delta_\xi\theta) =0,
\eea 
{using the commutation relations \eqref{Com1}}.} 
The term $a_\xi$ is the \emph{Lagrangian anomaly}, while  
$A_\xi$ will be referred to as the \emph{symplectic anomaly}.
For gravity we have formulations, such as Einstein--Hilbert or Einstein--Cartan, where 
both anomalies vanish. These are understandably the most studied formulations.
This means that in the gravity case all anomalies enters through the choice of a boundary Lagrangian which is needed  when dealing with boundaries and infinities.

In light of these considerations, it will be important  for us to decompose the component $\ell_\xi$ which enters the Lagrangian variation \eqref{deltal} as follows,
\be
\ell_\xi = \iota_\xi L + a_\xi. 
\ee
The first term on the RHS  is the usual expression that arises when the Lagrangian transforms covariantly 
under diffeomorphisms. 
Hence, for  a covariant Lagrangian  $a_\xi=0$. 
As shown in Appendix~\ref{Noetherapp}, the Noether charge in the presence of anomalies is given by
\be\label{charge0}
Q_\xi = \int_S q_\xi,\qquad \rd q_\xi\heq   I_\xi \theta - \iota_\xi L - a_\xi.
\ee
The hallmark of Noether theorems is not only that the Noether charges are conserved, thanks to $\rd j_\xi \heq 0$, but also that the Noether charge is the canonical generator of symmetry under certain conditions.
More precisely,  we need to consider the \emph{fundamental canonical relation}\footnote{ The contraction $I_\xi \Omega= \delta_{\xi} \ip \Omega $ is also sometimes denoted 
$\Omega( \delta_{\xi},\delta)$.}
\be\label{Flux0}
-I_\xi \Omega = \delta\left(\int_\Sigma C_\xi \right) +  \delta Q_{\xi} -\cF_\xi.
\ee
We refer to $\cF_\xi$ as the \emph{Noetherian flux}. It appears simply as the component of $I_\xi\Omega$ not contained in the total variation of the total Noether charge of diffeomorphism. The explicit  on-shell expression 
of the flux  is given in equation \eqref{Fluxdef}. Its off-shell expression is given in \eqref{offshell}.
The qualifier  Noetherian, for the flux,  indicates that the split between charge and flux on the RHS is not arbitrary\footnote{ This prescription is closer to the Wald--Zoupas approach \cite{Wald:1999wa} and should be contrasted with the  approach in  \cite{Barnich:2011mi, Barnich:2010eb, Compere:2018ylh, Compere:2020lrt}. In the latter,  there is no a priori prescription for the split between  the charge and the flux.
 This ambiguity is also  reflected
 in the Barnich--Troessaert bracket which yields the appearance of cocycles. As we are going to show in Section \ref{sec:IBra}, our Noetherian prescription allows us to define a charge bracket which represents the symmetry algebra  faithfully. } but associated with the Noether charge, which in turns is associated to a unique choice of Lagrangian as explained above.

 When $\cF_\xi=0$, we have that the Noether charge is the Hamiltonian charge, that is the canonical generator of symmetry. 
This happens, for instance, for  corner symmetries \cite{DonnellyFreidel, Freidel:2015gpa, Freidel:2016bxd, Freidel:2020xyx,  Freidel:2020svx, Freidel:2020ayo} which are  field-independent diffeomorphisms leaving the corner surface $S$ fixed.
It can also happen when there is a boundary and boundary conditions are imposed.
More precisely, when  there is no anomaly and  $\xi$ is \emph{field independent}, we have that the Noetherian flux is the symplectic flux $\cF_\xi=\cF_\xi^{\theta}$. 
In this case, imposing boundary conditions that requiring that    no symplectic flux leaks through the boundary also implies  that the Noetherian flux vanishes.
These points are developed in Section \ref{sec:NvH}.

\subsection{Noetherian  flux}\la{sec:Ano}

In this section we investigate the structure of the Noetherian flux $\cF_\xi$.
It is important to appreciate that this flux is, like the symplectic potential $\theta$ and the Noether charge $Q_\xi$, \emph{uniquely} determined by the choice of Lagrangian.
It is supported, on-shell, at the corner and  given by 
\be\label{Fluxdef}
\cF_\xi \heq\int_S  (\iota_\xi \theta + q_{\delta \xi} + A_\xi)\,,
\ee
that is the sum of three terms. The first term is the symplectic flux given by the contraction $\iota_\xi\theta$. As we have seen, this terms is due to the fact that there can be leak of symplectic flux through the motion of the boundary $S$ when it is translated along $\xi$.
The second term appears if the vector field $\xi$ is field dependent, 
while the third term appears if there are symplectic anomalies.
In the usual case of field-independent transformations and a choice of symplectic potential with no anomaly, such as Einstein--Hilbert or Einstein--Cartan formulations, only the first term contributes. 

The proof of \eqref{Fluxdef} goes as follows. One first evaluates the symplectic anomaly from first principle using  repeatedly the  equations $I_\xi \theta \heq  \rd q_{\xi} +\iota_\xi L + a_\xi$ and $\delta L\heq \rd\theta $, valid on-shell. 
One gets 
\bea
\Delta_\xi\theta&=& 
(\delta_\xi -{\cal L}_\xi- I_{\delta \xi})\theta \cr
&=& I_\xi \delta \theta + \delta I_\xi\theta   -\iota_\xi \rd \theta - \rd \iota_\xi\theta - I_{\delta \xi}\theta \cr
&\hat{=}& I_\xi \delta \theta + \delta (I_\xi\theta- \iota_\xi L)  - (I_{\delta \xi}\theta- \iota_{\d\xi} L)
 - \rd \iota_\xi\theta \cr
& \hat{=}& I_\xi \delta \theta + \delta (\rd q_\xi + a_\xi )  - (\rd q_{\delta \xi} + a_{\delta \xi})  -\rd \iota_\xi \theta \cr
&{=}&  I_\xi \delta \theta +\delta a_\xi -a_{\delta\xi} + \rd \left[\delta q_\xi - (\iota_\xi \theta + q_{\delta \xi})\right]\,.\la{DT}
\eea
One can then use the definition \eqref{covarianceprop} of the symplectic anomaly $A_\xi$ to rewrite this identity as 
\be 
- I_\xi \delta \theta \heq  \rd \left[\delta q_\xi - (\iota_\xi \theta + q_{\delta \xi}+ A_\xi)\right],
\ee 
which gives us (\ref{Flux0}), (\ref{Fluxdef})   after integration over a slice $\Sigma$ with boundary $S$.

Note that the flux expression \eqref{Fluxdef} can be used to give a more elaborate proof of the symplectic flux equation \eqref{sympflux}.
This is obtained as follows,
\bea \pa_t \Omega_t& =& 
\int_\Sigma \cL_\xi \omega
=\int_\Sigma \left( \d_\xi -\Delta_\xi -I_{\d\xi}\right)\omega
\cr
&\, =\,&  \delta  \cF_\xi
-\int_\Sigma (\Delta_\xi +I_{\d\xi}) \omega\cr
&=&
\delta  \cF_\xi
-\int_\Sigma I_{\d\xi} \omega
-\int_\Sigma( \d\Delta_\xi \theta-\Delta_{\d\xi}\theta )\cr
&\heq&
\delta  \left(\cF_\xi - Q_{\d\xi}  -\int_S A_\xi \right)  - \left( \cF_{\d\xi}-\int_S A_{\delta \xi} \right)\cr
&\heq& \delta \cF^\theta_\xi -\cF^\theta_{\delta \xi}\,,
\eea 
where in the first line we have used the definition of the anomaly \eqref{ano},
in the second line $\int_\Sigma \d_\xi \omega=\d\int_\Sigma  I_\xi \omega=
\delta  \cF_\xi$ due to the fundamental canonical relation  \eqref{Flux0}, in the third the commutation relation \eqref{Com1}, in the fourth line
we have used the fundamental canonical relation  \eqref{Flux0}  for a degree-one (or ``fermionic'') vector field $\delta\xi$\footnote{Beware of the sign changes when the vector field is a field-space one-form, 
\be
-\int_\Si I_{\d\xi}\om = -\d Q_{\d\xi} -{\cal F}_{\d\xi}\,, \quad \Delta_{\d\xi} \theta = - \delta a_{\d\xi} + \rd \d A_\xi\,,
\ee
whereas $\delta \Delta_\xi \theta =-\delta a_{\delta \xi} + \delta \rd A_\xi$.
}
and the symplectic anomaly  \eqref{covarianceprop} to establish that
\be
\delta \Delta_\xi \theta - \Delta_{\delta \xi} \theta = \delta \rd A_\xi-\rd A_{\delta \xi}\,.
\ee
This alternative proof shows explicitly how the Lagrangian anomaly drops out and does not contribute.

\subsection{Changing the Noetherian split}
In the previous sections we have performed a decomposition of the symplectic contraction in terms of an integrable component given by the  Noether charge $Q_\xi$ and a flux component
$\cF_\xi$. This decomposition depends on the choice $(L,\theta)$. It is therefore natural to wonder how this decomposition changes under a change of boundary Lagrangian
$L'= L+\rd \ell$ with associated symplectic potential $\theta'$ defined in \eqref{L'}.
To be general, we will not assume that the boundary Lagrangian is covariant.
The Lagrangian and symplectic anomalies transform as follows,
\be
a'_\xi = a_\xi + \Delta_\xi \ell, \qquad 
A'_\xi= A_\xi -  \Delta_\xi \vartheta\,.
\la{an-shift}
\ee
{The first shift follows immediately from \eqref{Com1} and the first relation in \eqref{covarianceprop}; the second follows from the expression 
\eqref{L'} for the shifted potential, using \eqref{Com1},  the second relation in \eqref{covarianceprop}, {and the first shift}.
}

So we see that even if we start with a fully covariant formulation where $a_\xi=A_\xi=0$, 
boundary shifts can create anomalies.
The new charges and fluxes are 
\be\la{Q'}
Q'_\xi =  \int_\Sigma \left(I_\xi\theta' -\iota_\xi L'-a'_\xi\right),
\qquad
\cF'_\xi= \int_S \left(\iota_\xi \theta' + q'_{\delta \xi} +  A'_\xi \right).
\ee
We show in Appendix \ref{AppD} that these are related to the old ones by the following shifts
\be \label{trans1}
Q'_\xi-Q_\xi =  \int_S (\iota_\xi \ell-I_\xi \vartheta ),
\qquad 
\cF'_\xi-\cF_\xi 
=  \int_S \left(\delta \iota_\xi \ell-\delta_\xi \vartheta \right) .
\ee
These shifts of charge and flux preserve the fundamental canonical relation \eqref{Flux0}, since the symplectic form is shifted by the corner potential as in \eqref{shiftO}.
What is remarkable in these formulae is the fact that all the anomaly contributions have finally dropped out of the shifts. In particular, notice that as a consequence of \eqref{trans1}, if one starts with a covariant bulk Lagrangian, the final
expression for the charge is anomaly-free even if the boundary
Lagrangian is anomalous.

\section{Charge bracket}\label{sec:3}

The previous section introduced the notion of Noetherian split in order to clearly distinguish what we call charge from what we call flux.
It is also convenient to define the notion of an equivalence $L \sim L+\rd \ell$ when two Lagrangians are related by a boundary shift. We denote such a class of Lagrangians by $[L]=[L+\rd \ell]$. 
The equivalence class $[L]$ characterizes the equations of motion. We will say that a class is \emph{covariant} when there exists a representative $L_c$ of the class with no anomaly. 
For gravity we have that the class is covariant with covariant representative the Einstein--Hilbert Lagrangian.
The choice of Lagrangian representative $L$ within an equivalence class  can be related, as we will see in Section \ref{sec:NvH}, to a choice of boundary condition.
The choice of Lagrangian $L$ also determines  uniquely, as we have seen, the Noetherian charges and fluxes. 

 In this section we show that it is possible to define a charge bracket associated with a given Lagrangian that takes into account the presence of a non-vanishing flux. This bracket defines flux-balance laws which encode the dynamics of the theory, and is used to  provide a canonical representation of the extended corner symmetry algebra derived below.

\subsection{Brackets and symmetric flux-balance law}

We are interested in constructing the bracket in the more general framework that includes field-dependent symmetry generators as well as anomalies, since both features arise commonly when dealing with boundaries.
More precisely, we have emphasized in  Section \ref{sec:Ano} that  the equality $\delta_\xi = {\cal L}_\xi$ is not always satisfied in the presence of background structures which create Lagrangian and symplectic anomalies.
Another instance where the equality is also not satisfied is when we allow the vector fields to be 
field dependent. In these cases, anomalies appear in the construction of the symmetry bracket. The field dependency of the generators  also affects the construction of the Noetherian flux as shown in \eqref{Fluxdef}.
The presence of field dependent vector fields is \emph{unavoidable} when we work in a fixed gauge. 
The introduction of field dependent vector fields $\xi$ is also an integral part of the construction of an extended phase space that includes edge modes \cite{DonnellyFreidel,Freidel:2020xyx,Freidel:2020svx,Freidel:2020ayo}.
Finally it is a central part in the recent construction of \cite{Adami:2020ugu, Ruzziconi:2020wrb}, which involves a field dependent redefinition of the symmetry generators in order  to reabsorb fluxes.

 By construction the commutator of two symmetry transformations is itself a symmetry transformation.
This means that there exists a bracket $\lbr \cdot ,\cdot \rbr $ such that
 \be\label{maincom}
 [ \delta_\xi,  \delta_\chi]= -\delta_{ \lbr \xi,\chi\rbr}\,.
 \ee
 For field independent transformations the bracket is simply the Lie bracket.
 In general, however, we have that the bracket is given by \cite{Barnich:2010eb}\footnote{ The proof follows from
 \bea
 \delta_\xi \delta_\chi g_{\m\n} =
  \delta_\xi ({\cal L}_\chi g_{\m\n})=  {\cal L}_{\delta_\xi\chi} g_{\m\n}
  + {\cal L}_\chi (\delta_\xi g_{\m\n})= {\cal L}_{\delta_\xi\chi} g_{\m\n}  + {\cal L}_\chi {\cal L}_\xi g_{\m\n}. 
 \eea
 Antisymmetrizing this equation and using 
 $[{\cal L}_\chi, {\cal L}_\xi] = {\cal L}_{ [\chi, \xi]_{\mathrm{Lie}}} $ we obtain \eqref{maincom}.
}
 \be\la{vbra}
  \lbr\xi,\chi\rbr:= [\xi, \chi]_{\mathrm{Lie}} + \delta_\chi \xi - \delta_\xi \chi.
 \ee
This bracket also enters 
\be\la{DI}
[\d_\xi, I_{\chi}]=-I_{\lbr\xi, \chi\rbr}\,,
\ee
as well as the commutation relations of the anomaly operator
\be\label{anomaly}
[\Delta_\xi ,\Delta_\chi]= -\Delta_{\lbr\xi,\chi\rbr}, \qquad [\Delta_\xi, I_{\chi}]=I_{\delta_\chi\xi} -I_{\lbr\xi,\chi\rbr}, \qquad
[\Delta_\xi, \iota_{\chi}]=\iota_{\delta_\chi\xi} -\iota_{\lbr\xi,\chi\rbr}.
\ee
These identities are proven in Appendix \ref{CartanC}.

If one starts with the fundamental equation  \eqref{Flux0}  and contracts it 
  with a field variation $\d_\chi$ we get the equation
  \be \la{IIO}
\Omega(\delta_\xi, \delta_\chi):= I_\chi I_\xi \Omega \heq  \delta_\xi Q_{\chi}- I_\xi\cF_{\chi}. 
\ee
It is important to appreciate that the freedom of choice between charge and flux is restricted by demanding that the charge and the flux are \emph{Noetherian} as in \eqref{charge0}, \eqref{Fluxdef}. This  means that they descend from a symplectic potential $\theta$ and a Lagrangian $L$, as explained above.

Now we can present the {\it symmetric flux-balance relation}.
Since the symplectic structure is antisymmetric, we have that 
$\Omega(\delta_\xi,\delta_\chi)=- \Omega(\delta_\chi,\delta_\xi)$, which in turn  implies that
we have the on-shell equality
\be\label{Balance1}
 \delta_\xi Q_\chi - I_\chi \cF_\xi  \heq - (\delta_\chi Q_\xi - I_\xi \cF_\chi )\,  .
\ee
Two facts are remarkable about this identity. First, it  is independent of the split between
charge and flux  and therefore invariant under asymptotic renormalization. 
Indeed, under a change of boundary Lagrangian, we have that the LHS of \eqref{Balance1}  transforms as 
\be
 \delta_\xi (\iota_\chi \ell-I_\chi \vartheta )
- I_\chi (\delta \iota_\xi \ell-\delta_\xi \vartheta )
= 2\delta_{[\xi} \iota_{\chi]} \ell 
+I_{\lbr\xi,\chi\rbr}\vartheta,
\ee
which is manifestly antisymmetric and is equivalent to the skew-symmetry of the bracket.
Second, the symmetric flux-balance relation is obtained by using the  field equations.
What is remarkable is that the reverse is also true:
given the flux and the symplectic potential, the kinematical flux-balance law \eqref{Balance1}
 \emph{implies} the validity of the Einstein's equations selected by the symmetry vectors, as shown below in \eqref{Braoff}. This has already been verified at asymptotic null infinity  in \cite{Freidel:2021yqe}.
   
   \subsection{An invariant  Poisson bracket}\la{sec:IBra}
 While the symmetric flux-balance relation \eqref{Balance1} provides a very useful and powerful tool to recover some of the Einstein's equations on $S$, it does not provide  access to the structure of the symmetry group, as it is not written in terms of a bracket providing a representation of the  symmetry algebra on the corner sphere. We will now show how the Noetherian split we advocate for allows us to go one step further
 and define a second flux-balance relation in terms of a bracket available for open Hamiltonian systems and in the presence of anomalies.
 
 Before doing so, we need to establish a technical lemma.
 Let us assume that a $d$-dimensional spacetime $M$ is such that $H_{d-1}(M)=0$.
 Then there exists a two form $c_{(\xi,\chi)}$ which is such
 \be\label{c}
\rd c_{(\xi,\chi)} =   \Delta_\xi a_\chi -\Delta_\chi a_\xi + a_{\lbr \xi,\chi\rbr}.
 \ee
 Moreover, $c_{(\xi,\chi)}$ is independent of the choice of representative within a class $[L]$.

To prove this, one first notices  that the RHS of \eqref{c} is a closed 3-form, which follows from the definition \eqref{covarianceprop} of the Lagrangian anomaly,  
  \be
  \rd (\Delta_\xi a_\chi -\Delta_\chi a_\xi + a_{\lbr \xi,\chi\rbr})=
  ([\Delta_\xi,\Delta_\chi]+ \Delta_{\lbr \xi,\chi\rbr})L=0.
  \ee
 The fact that the 2-form $c$ is independent of the choice of boundary Lagrangian follows from the shift 
 \eqref{an-shift} of the Lagrangian anomaly, namely  $a'_\xi = a_\xi + \Delta_\xi\ell$, from which
 \be
 c'_{(\xi,\chi)} = c_{(\xi,\chi)} + [\Delta_\xi,\Delta_{\chi}] \ell + \Delta_{\lbr \xi,\chi\rbr}\ell = c_{(\xi,\chi)}.
 \ee
 Hence, if the class $[L]$ admits
 a representative Lagrangian  free of anomaly, like in the case of gravity,\footnote{ Note that if we have a non-zero cosmological constant then the Einstein--Hilbert Lagrangian is infinite on-shell. If we restrict the admissible class $[L]_{\mathrm{finite}}$ of Lagrangians to the ones that are finite on-shell, we will have a residual anomaly $c_{(\xi,\chi)}$ associated to  the equivalence class,  which could be the source of Brown--Henneaux central charges \cite{Brown:1986nw}.} we have 
 $c_{(\xi,\chi)}=0 $. This simplifies the definition of the bracket, as we are about to see.

 We are now in a position to define the following corner symmetry bracket suggested by the identity  \eqref{Balance1}:
 \be\label{IBra}
  \boxed{\,\,
  \{Q_\xi,Q_\chi\}_{L}:= \delta_\xi Q_\chi - I_\chi \cF_\xi
   + K^L_{(\xi,\chi)}
  \,,\,\,}
 \ee 
 where $K^L_{(\xi,\chi)}$ is given by
 \be
 K^L_{(\xi,\chi)} := \int_S \iota_\xi \iota_\chi L +\int_S  (\iota_{\xi}  a_\chi - \iota_\chi  a_\xi )
  + \int_S c_{(\xi,\chi)}\,,
 \ee 
 and it reduces simply to the Lagrangian contraction term when there is no anomaly.
 This bracket, whose label $L$ is meant to stress its dependence on the Lagrangian and not just Lagrangian class,  represents a generalization 
  of the  Barnich--Troessaert bracket \cite{Barnich:2009se,  Barnich:2010eb, Barnich:2011mi,Troessaert:2015nia}, in that it is modified by the Lagrangian term, and it is extended to include the case of anomalies and all normal translations. It also generalizes the work of \cite{Chandrasekaran:2020wwn} to the case of field dependent vector fields.
  The Lagrangian term is non-zero only when we consider vector fields $(\xi, \chi)$ which are transverse to $S$ and form a 2-dimensional basis of the normal bundle $(TS)^{\perp}$.
  The relevance  of such a term was first noticed by Speranza in \cite{Speranza:2017gxd}.
  The symmetric flux-balance relation is simply the statement that this bracket is antisymmetric since $K^L_{(\xi,\chi)}$ is manifestly skew.

  The bracket  \eqref{IBra} satisfies {two} essential properties:
 \begin{enumerate}
 \item It satisfies the Jacobi identity,
\item It provides a \emph{representation} of the commutator \eqref{maincom} for any 
$L$ in the class $[L]$.
 \end{enumerate}

The second property implies the first one, so we focus on the second property first. To  prove it one  computes, using  the definition  \eqref{charge0} of $Q_\chi$ and the commutation relations \eqref{anomaly},  the charge anomaly:
 \be\label{QAnomaly}
\Delta_\xi Q_\chi=
Q_{\delta_\chi\xi}
-Q_{\lbr\xi,\chi\rbr}
+\int_S  (I_\chi  A_\xi + i_\chi  a_\xi )
- \int_S c_{(\xi,\chi)}\,,
\ee
where we have used \eqref{c}.
 We can now get the flux-balance equation  from 
 \bea
 I_\chi \cF_\xi - \int_S  i_{\xi} i_{\chi}L
 &\heq&  \int_S \left(i_{\xi}  (I_\chi {\theta}- i_{\chi}L) + q_{\delta_\chi \xi} + I_\chi A_\xi\right)\cr
   &=&  \int_S \left(\cL_{\xi}  q_\chi +i_{\xi}  a_\chi + q_{\delta_\chi \xi} + I_\chi A_\xi \right)\cr
    &=& \d_\xi Q_\chi -\Delta_\xi Q_\chi + Q_{\delta_\chi \xi}
    +  \int_S \left(i_{\xi}  a_\chi  + I_\chi A_\xi \right)\cr
    &=& \d_\xi Q_\chi
    + Q_{\lbr\xi,\chi\rbr}
+\int_S  (i_{\xi}  a_\chi - i_\chi  a_\xi +c_{(\xi,\chi)})\,.
 \eea
Plugging this equality in \eqref{IBra} shows 
that the bracket provides, 
even in the presence of anomalies,  the
{\it fundamental charge commutation relation}
\be\label{Flux1}
\boxed{\,\,
\{ Q_\xi, Q_\chi\}_{L} \heq -Q_{\lbr\xi,\chi\rbr}\,.\,\,
}
\ee

The second aspect of the second property  means that the charge bracket relation \eqref{Flux1} is independent  of the split between charges and fluxes.
 To see this, let us consider how this bracket transforms under a change of Lagrangian $L'=L+\rd\ell$ and 
  symplectic potential $\delta \theta'= \delta \theta - \rd\d \vartheta$. 
  We have already seen that the 2-form $c_{(\xi,\chi)}$ is independent of the choice of boundary Lagrangian.  Then, from the transformations \eqref{trans1}, we can evaluate that the new bracket is related to the old one as follows:\footnote{We use the commutator \eqref{DI} and that 
 \be
\iota_{\chi} \cL_\xi  \ell -\iota_\xi \cL_{\chi} \ell +
\iota_{\delta_\xi \chi}\ell -\iota_{\delta_\chi \xi}\ell =
-\iota_{\lbr \xi,\chi\rbr}\ell -\iota_\xi \iota_\chi \rd \ell
+\rd(i_\xi i_\chi \ell)\,.\la{idL}
\ee
}
 \begin{align}
  \{Q'_\xi,Q'_\chi\}_{L'}&= \{Q_\xi,Q_\chi\}_{L} + 
  \int_S (\delta_\xi  \iota_\chi \ell- \delta_\xi I_\chi \vartheta ) 
  -  \int_S \left( \delta_\chi \iota_\xi \ell-  I_\chi \delta_\xi \vartheta \right) + \int_S \iota_\xi \iota_\chi \rd \ell
  \cr
  &\quad+ \int_S \left( \iota_\xi \Delta_{\chi} \ell -\iota_{\chi} \Delta_\xi  \ell \right)\\
  &= \{Q_\xi,Q_\chi\}_{L} +  \int_S (I_{\lbr \xi,\chi\rbr}\vartheta - \iota_{\lbr \xi,\chi\rbr}\ell)\,,
  \label{brack2}
 \end{align}
from which we finally get
 \be\label{brack3}
  \{Q'_\xi,Q'_\chi\}_{L'} + Q'_{\lbr \xi,\chi\rbr} = 
   \{Q_\xi,Q_\chi\}_{L} + Q_{\lbr \xi,\chi\rbr} \,.
 \ee
 This  relationship shows that the canonical relation  \eqref{Flux1} is preserved under the change of boundary Lagrangian  and \eqref{Flux1} shows  
 that the Jacobi's identity is  always satisfied by the bracket \eqref{IBra}. 
 
 It is important to appreciate that the canonical commutation relation  can also be written as a  {\it fundamental flux-balance law}. Explicitly, if one uses the bracket and charges that come from a covariant Lagrangian, we see that  \eqref{Flux1} is equivalent to 
 \be
 \delta_\xi Q_\chi + Q_{\lbr\xi,\chi\rbr}\, \heq \,I_\chi \cF_\xi + \int_S  \iota_\chi \iota_\xi L.
 \ee
 This equation describes the change of $Q_\chi$ as it is moved along $\xi$.
 The first component of the motion is a rotation inside the symmetry orbit,  while the Noether flux and the contracted Lagrangian appear as  source terms which prevent the charge evolution to be purely along the coadjoint orbit.
 
 Finally, in  Appendix \ref{BracketS}  we  give the proof of the relation
 \be\la{Oxc}
 \Omega(\d_\xi, \d_\chi)\heq
 \d_\xi Q_\chi -\d_\chi Q_\xi
 +Q_{\lbr \xi, \chi\rbr} 
 +K^L_{(\xi,\chi)}. 
 \ee
 A direct derivation of the formula \eqref{Oxc} can be obtained by plugging  \eqref{IIO}
 into   \eqref{Flux1} with the definition \eqref{IBra}.

 As usual, the fact that the symplectic form is closed, i.e. $\delta \Omega=0$, is equivalent to the identity 
 \be \label{cyclic}
 \delta_{\xi} \Omega(\delta_\chi, \delta_\zeta) + \Omega(
 \delta_{\lbr \xi, \chi \rbr}, \delta_{ \zeta}) +\mathrm{cycl.} =0\,,
 \ee 
 where cycl. refers to the cyclic permutation of $(\xi,\chi,\eta)$.
 We can also see that the 
 charge dependent terms in \eqref{Oxc} drop out of \eqref{cyclic}, which therefore yields a cocycle identity for the  contribution  $K^L_{(\xi,\chi)}$, namely
\be 
 \delta_\xi K^L_{(\chi,\zeta)} + 
 K^L_{(\lbr \xi,\chi\rbr, \zeta) } + \mathrm{cycl.} =0.
 \ee This cocycle identity  for the anomaly terms inside $K^L_{(\xi,\chi)}$ was proven in \cite{Chandrasekaran:2020wwn}, for the case where  the Lagrangian $L$ belonged to a covariant class. That is
 when $a_\xi=\Delta_\xi \ell$ and $c_{(\xi,\chi)}=0$.
 Our proof generalizes this result.

  This suggests that we could have defined, following Barnich--Troessaert, a bracket that does not include the contribution from the Lagrangian and its anomaly:
  \be
  \{Q_\xi,Q_\chi\}_{L}^\mathsf{BT}:= \delta_\xi Q_\chi - I_\chi \cF_\xi. 
  \ee
  The relationship between the two brackets is through the Lagrangian  $2$-cocycle 
  \be 
  \{Q_\xi,Q_\chi\}_{L}^{\mathsf{BT}}
  = \{Q_\xi,Q_\chi\}_{L} -  K^L_{(\xi,\chi) }.
  \ee
  This shows that the cocycles attached to the  Barnich--Troessaert bracket in \cite{Compere:2020lrt}  can all be derived from the non-covariance of the boundary Lagrangian (an explicit example of this statement was given in \cite{Freidel:2021yqe}). 
  
It is not clear to us that the Barnich--Troessaert bracket $\{\cdot,\cdot\}_{L}^{\mathsf{BT}}$ satisfies the Jacobi identity, since this requires proving two key properties:
\be
\{Q_\zeta,K^L_{(\xi,\chi) } \}_{L}^{\mathsf{BT}}\stackrel{?}{=}\delta_\zeta K^L_{(\xi,\chi) },
\qquad
\{K^L_{(\xi,\chi) }, K^L_{(\xi',\chi') }\}_{L}^{\mathsf{BT}}\stackrel{?}{=}0.
\ee 
While the first identity is plausible, the second one seems far less obvious as $K^L_{(\xi,\chi) }$ is a \emph{field dependent} cocycle.
These identities are usually \emph{postulated} but, as far as we can tell, never proven.
A proof  of these identities would require showing that the Barnich--Troessaert bracket  can be extended to the entire gravitational phase space and not just to the subset of observables associated with symmetry charges.

 \subsection{Algebra and constraints} \la{sec:constraints}
 
What we have proven so far is the fact that the fundamental commutation relation \eqref{Flux1} is satisfied, provided we assume the validity of the bulk constraint equations.
 We now want to ask: what  happens if we do not assume that?
 The main claim of our work is that the demand of the validity of \eqref{Flux1}
 \emph{implies}  Einstein's equations.
 In other words, we are saying that the demand that the corner charges 
 form a faithful and centerless representation of the boundary symmetry algebra at any corner is the essence of Einstein's equations.

The first step of the argument requires the computation of the anomaly of the constraint $C_\xi$.
One first establishes that 
\be\label{DC}
C_{(\xi,\chi)} := \Delta_\xi C_\chi -
 C_{\delta_\chi  \xi} + C_{\lbr\xi,\chi\rbr}=0.
\ee
To see this, one uses the fact that the constraints are related to the equations of motion 
by $\rd C_\xi = I_\xi E$ and, since 
the equations of motion have no anomaly ($\Delta_\xi E=0$), one gets, using \eqref{anomaly}, that 
\bea
\rd (\Delta_\xi C_\chi)= \Delta_\xi I_\chi E = [\Delta_\xi, I_\chi] E 
= I_{\delta_\chi\xi}E  -I_{\lbr\xi,\chi\rbr}E
= \rd( C_{\delta_\chi\xi} - C_{\lbr\xi,\chi\rbr}).
\eea
This establishes that  $C_{(\xi,\chi)}  =\rd \rho_{(\xi,\chi)} $,
 where $\rho_{(\xi,\chi)}$ is a codimension-$2$ form which depends linearly on $\xi $ and $\chi$.
We can  integrate this identity on $\Sigma$ and  use that $\xi$ and $\chi$ can be chosen 
 arbitrarily in the bulk of  $\Sigma$ while they can be chosen to  vanish on $S$. This means that the bulk and the boundary identities need to be satisfied independently, \emph{i.e.} one has 
 $C_{(\xi,\chi)}=0= \rd \rho_{(\xi,\chi)}$.
 One can also establish the identity \eqref{DC},  in a pedestrian manner, by using the explicit expression 
 \be\la{Ccon}
 C_\chi = \chi^\m G_{\m}{}^\n \epsilon_\n, \qquad \epsilon_\m = \iota_{\pa_\m}\epsilon,
 \ee
 where $\epsilon$ is the 4-volume form and $G_{\m\n}$ is the Einstein tensor.
 One has 
 \be
 \Delta_\xi  C_\chi  = (\Delta_\xi\chi)^\m G_{\m}{}^\n \epsilon_\n
 = C_{\delta_\xi \chi}  - C_{[\xi,\chi]_{\mathrm{Lie}}},
 \ee
 which is equivalent to \eqref{DC}.
 One now has to evaluate the form $\rho_{(\xi,\chi)}$ explicitly.
 We find that the identity 
 $\rd \rho_{(\xi,\chi)}=0$ gives, after integration over the hypersurface $\Sigma$, the remarkable off-shell bracket identity:
 \be\la{Braoff}
 \boxed{\,\,  \{ Q_\xi, Q_\chi\}_L+Q_{\lbr\xi,\chi\rbr}+ \int_S\iota_\xi  C_\chi =0.\,\,}
 \ee
 To establish this relation, we use that the dependency on the constraints can be restored through the shifts of the charge and flux
 \be\label{offshell}
 Q_\chi \rightarrow  Q_\chi  +\int_\Sigma   C_\chi\,,\qquad
 \cF_\xi \rightarrow  \cF_\xi + \int_\Sigma (C_{\delta_\chi \xi} + \iota_\xi  E).
\ee
The shift for the charge goes back to the original definition  \eqref{Noether}.
The fact that there is a bulk contribution to the Noetherian flux is surprising and often overlooked.
This bulk contribution vanishes {off-shell} when $\xi$ is field independent and $\xi$ is tangent\footnote{ If $\xi $ is tangent to $\Sigma$, then $\iota_\xi  E$ is a three form that integrates to zero when pulled back on $\Sigma$.} to $\Sigma$. It is thus only relevant  for the Hamiltonian constraint,  that is when considering vector fields $\xi$ which are transverse to the slice $\Sigma$. This shows that the Hamiltonian constraint is in fact Hamiltonian only on-shell of the equations of motion!

These shifts imply in turn that the combination entering the definition of the bracket bracket is given by 
\be
 (\d_\xi Q_\chi - I_\chi \cF_\xi) \rightarrow  (\d_\xi Q_\chi - I_\chi \cF_\xi) 
 -\int_S   \iota_\xi C_\chi 
 +\int_\Sigma ( \Delta_\xi  C_\chi - C_{\delta_\chi \xi} )\,,
\ee
from which we get 
\be
 \{ Q_\xi, Q_\chi\}_L+Q_{\lbr\xi,\chi\rbr}+ \int_S   \iota_\xi C_\chi =-\int_\Sigma ( \Delta_\xi  C_\chi - C_{\delta_\chi \xi} + C_{\lbr\xi,\chi\rbr})=0,
\ee
as promised.
More details about these calculations are given  in the appendix \ref{Mainf}.

\subsection{Noether versus Hamiltonian charges}\la{sec:NvH}
We have constructed a canonical algebra for the Noether charges in the general case of an open system with non-zero symplectic flux.
There are two instances however where we already know what the bracket of charge is.
The first instance is when the system is \emph{Hamiltonian}, i.e. admits no symplectic flux and $\cF^{\th}_\xi=0$, and  the second is when the  symmetry transformations are \emph{integrable}, i.e. $\cF^{\th}_\xi$ is $\d$-exact.
In both cases we can define a phase space on which the charges act as canonical transformations and the bracket is simply the canonical Poisson bracket obtained by inverting the conserved symplectic form.
The goal of this section is to show that we can  recover from the bracket \eqref{IBra} the usual canonical Poisson bracket of Hamiltonian charges in both cases.

In both cases we will need to assume that $\delta \xi=0$. The prototypical example of this recovery,
studied in detail by Harlow and Wu in \cite{Harlow:2019yfa}, is the case of gravity with \emph{Dirichlet} boundary conditions $\delta \bar{g}_{ab}=0$, where $\bar{g}_{ab}$ is the metric induced on a timelike surface. In this case, it is well known \cite{Iyer:1994ys} that the  Einstein--Hilbert Noether charge,   associated with a diffeomorphism that translates the boundary along itself, and also known as the Komar charge, is not the Hamiltonian charge.
We know, however, that the system is Hamiltonian, in  the sense of a conserved symplectic form, due to the Dirichlet boundary conditions. This means that we can define an Hamiltonian generator which  differs form the Einstein--Hilbert Noether charge.
This Hamiltonian generator is the Brown--York generator.
In \cite{ Freidel:2020xyx} it is shown that the difference between the Komar and the Brown--York charges  comes entirely from the presence of a boundary Lagrangian given by the Gibbons--Hawking term \cite{Gibbons:1976ue}.
This proves that the Hamiltonian charges associated with Dirichlet boundary conditions are in fact \emph{Noether charges} for the Gibbons--Hawking Lagrangian\footnote{ Which is the Einstein--Hilbert Lagrangian augmented by the Gibbons--Hawking term. Explicitly,
\be 
L_{\mathsf{GH}}=\tfrac12 R - \nabla_\mu(s^\mu K + \nabla_s s^\mu)\,,
\ee 
with $s^\mu$ a unit spacelike normal.} and that the canonical bracket for the 
Dirichlet Hamiltonian charges  is the same as the  Gibbons--Hawking bracket $\{\cdot, \cdot\}_{L_{\mathsf{GH}}}$.

This phenomenon is absolutely general.
Given a Lagrangian $L$ within a class $[L]$ we
construct a symplectic potential current  $\theta^L$, a symplectic form current $\omega^L$, and a collection of charges $Q^L_\xi$.
Most importantly, and as  explained in detail in \cite{Freidel:2020xyx}, we can also associate to the Lagrangian a boundary condition  ${\cal B}^L$ given by 
\be \label{BCL}
{\cal B}^L: \qquad \theta^L \stackrel{\Gamma}{=}0\,,
\ee where  $\Gamma$ is a time-like or null boundary with normal $s_\mu$ and $\stackrel{\Gamma}{=}$ means that the equality is considered as a pull-back of forms on $\Gamma$.
For instance,
since the symplectic potential of Einstein--Hilbert 
is $\theta_{\mathsf{EH}}\stackrel{\Gamma}= \tfrac{\sqrt{g}}2 (g^{\alpha \beta} s_\mu - s^\alpha \delta_\mu^\beta )\delta \Gamma_{\alpha \beta}^\mu$,
the Einstein--Hilbert boundary condition is of Neumann type, which imposes certain connection coefficient to be fixed on $\Gamma$. While the Gibbons--Hawking symplectic symplectic potential is 
$\theta_{\mathsf{GH}}\stackrel{\Gamma}= \tfrac{\sqrt{g}}2 (K \bar{g}^{\alpha \beta} - K^{\alpha \beta}  )\delta \bar{g}_{\alpha \beta}$, with boundary condition $\delta \bar g_{\alpha\beta}=0$.
Another metrical example is studied in \cite{Oliveri:2019gvm} and it is given by the Lagrangian 
$L_{\mathsf{mixed}}= \tfrac12 R - \tfrac23 \nabla_\mu (s^\mu K+ \nabla_s s^\mu)$ which imposes mixed boundary conditions $\delta K=0= \delta \tilde{\bar g}_{ab}$, where $\tilde{\bar g}_{ab}$ is the unimodular metric.

In any case, the main point  is that there exists a correspondence $L\to {\cal B}^L$ between Lagrangians and boundary conditions via \eqref{BCL}.
Once the boundary condition ${\cal B}^L $ is imposed,
the symplectic form $\Omega^L_{\Sigma_t} $ for slices $\Sigma_t$ that intersect $\Gamma$ at $S_t$, is conserved.
This means that we have a phase space structure ${\cal P}^L$ and that we can construct the canonical Poisson bracket  
$\{\cdot,\cdot\}_L^{\mathsf{can}}$ on ${\cal P}^L$ by inverting the symplectic structure $\Omega^L_{\Sigma_t} $. 
It also means that we have a \emph{ Hamiltonian} generator $H_\xi^L$  on ${\cal P}^L$ which is such that $-I_\xi \Omega^L_{\Sigma}= \delta H^L_\xi$.

It is now natural to ask what is the relationship between our bracket $\{\cdot,\cdot\}_L$ and the canonical bracket $\{\cdot,\cdot\}_L^{\mathsf{can}}$? What is the relationship between the Noether charge $Q^L_\xi$  and the Hamiltonian charge $H^L_\xi$?
The answer is very simple, at least in the absence of anomalies. If we assume that the symmetries $\xi$ that preserve\footnote{This is the case for the diffeomorphisms tangent to $\Gamma$.} ${\cal B}_L$ also satisfy
$a_\xi=A_\xi=0$, then 
\be\label{canQ}
\{\cdot,\cdot\}_L = \{\cdot,\cdot\}_L^{\mathsf{can}}, \qquad 
Q^L_\xi = H^L_\xi, 
\ee
where the equality is valid when the boundary condition ${\cal B}^L$ is imposed.
See also~\cite{Margalef-Bentabol:2020teu} for a discussion of the relation between Noether and Hamiltonian charges with anomalies included.

To prove this, let us consider  a timelike or null boundary $\Gamma$ and let us denote $S_t=\Sigma_t\cap \Gamma$ the corner intersections at time $t$.
Let us assume that we have a covariant class $[L]$ and that    $L$ is the  covariant Lagrangian representative of the class $[L]$.
As already pointed out in \eqref{sympflux}, the demand of having a conserved symplectic form, 
 is equivalent to the vanishing of the symplectic flux  at the  corners $S_t=\Sigma_t\cap \Gamma$, namely
 \be\la{Ham2}
 \delta \cF^{\theta^L}_\xi=0\,.
 \ee
 Since $\cF^{\theta^L}_\xi =\int_{S_t} \iota_\xi \theta^L $, imposing \eqref{Ham2} for all vectors $\xi$ tangent to $\Gamma$ 
 means that there exists a boundary Lagrangian $\ell$ and a corner potential $\vartheta$ such that 
 \be
 (\theta^L+\d\ell)\stackrel{\Gamma}{=}\rd \vartheta\,. 
\ee
This is the result  of \cite{Harlow:2019yfa}, which framed in our context can be understood as fact that the symplectic potential $\theta^{L'}= \theta^L +\delta \ell -\rd \vartheta $ associated with the new Lagrangian 
$L'= L+\rd \ell$ satisfies the condition \eqref{BCL}, that is  $\theta^{L'}\stackrel{\Gamma}{=}0.$
This means that we can rewrite the condition \eqref{Ham2} as the condition that the symplectic flux associated with the new Lagrangian vanishes
\be \la{bcF}
{\cal B}^{L'}: \quad \cF^{\theta^{L'}}_\xi=0.
\ee
In the case of interest where we start from a covariant boundary Lagrangian and we assume that the pair $(\ell,\vartheta)$ defining ${\cal B}_L$  possess no anomaly under the  boundary symmetries,\footnote{This is the case for the Gibbons--Hawking boundary term which is covariant under diffeomorphism tangent to $\Gamma$.} we have that the Hamiltonian flux vanishes
$\cF^{{L'}}=\cF^{\theta^{L'}}=0$. We also have that the cocycle $K^{L'}_{(\xi,\chi)}$ vanishes, as
 for vector fields $(\xi,\chi)$ tangent to $\Gamma$ the contraction
$\int_S \iota_\xi\iota_\chi L =0.$
From \eqref{Flux0} and  \eqref{IBra} this means that 
\be
-I_\xi \Omega =\delta Q^L_\xi,
\qquad
\{ Q^L_\xi, Q^L_\chi \}_L= \delta_\xi Q^L_\chi = - Q_{\lbr\xi,\chi \rbr}.
\ee 
This establishes that $Q^L_\xi$ is the Hamiltonian charge and the bracket the canonical bracket.\footnote{The double-Lie bracket in the last equality can also be replaced with the basic Lie bracket since we are only considering field-independent diffeomorphisms. 
}

The case where the boundary Lagrangian is anomalous under the boundary symmetries is more involved and will be investigated elsewhere. What we can say is that the  preservation of the boundary condition $\cB^{L'}$ means  that $\delta_\xi \theta^{L'}=0= {\cL}_\xi\theta^{L'}$ which implies, given  \eqref{covarianceprop}, that the vector $\xi$ satisfies the condition
\be 
\delta a'_\xi + \rd A'_\xi =0.
\ee 
In this case, we expect the bracket $\{\cdot,\cdot\}_L$ to be related to the canonical bracket with the addition of central charges produced by the boundary anomalies. This is what happens in the asymptotic AdS case  \cite{Brown:1986nw}  and has also be demonstrated for the null case in \cite{Chandrasekaran:2020wwn}.

A third possibility to obtain integrable Noether charges is to achieve the condition $\cF_\xi=0$  by a choice of   ``slicing'' \cite{Adami:2020ugu, Ruzziconi:2020wrb}, i.e. a  choice of  field dependency for the vector fields $\xi$. A direct way to see how this can work is to recall the expression \eqref{Fluxdef} for the Noetherian flux. By a judicious choice of field-dependent diffeomorphism generators, one can seek to  fine-tune the second term  $q_{\d\xi}$ to cancel the other two (or simply the symplectic potential contribution in the case where this has no anomaly), namely to impose
\be
\int_S q_{\d\xi}= - \int_S (\iota_\xi \theta  + A_\xi)\,.
\ee
This procedure has been shown to yield integrable charges  in 2d and 3d gravity theories  \cite{Adami:2020ugu, Ruzziconi:2020wrb}, where there are no local propagating degrees of freedom.

\section{Extended corner symmetry
}\label{sec:4}

We now explicitly derive the form of the Noether charges of the corner symmetry algebra extension in the Einstein--Hilbert gravity formulation and compare with the study of null infinity performed in \cite{Freidel:2021yqe}.

\subsection{Extended corner symmetry algebra} \la{sec:extal}

As first revealed in \cite{DonnellyFreidel},  the gravitational symmetry revealed by the presence of a corner decomposes into the sum of surface diffeomorphisms, surface boosts and surface translations.
The \emph{corner symmetry group} is the subgroup of transformations that do not move the surface. It comprises of the surface diffeomorphism and surface boosts and has   the semidirect sum structure 
\be
\mathfrak{g}_S = \mathrm{diff}(S)\oright  \mathfrak{sl(2,\mathbb{R})}^S\,.
\ee
As shown in \cite{Donnelly:2020xgu}, this algebra appears as the Automorphism group of the normal bundle associated with the embedding of $S$ into spacetime.
The surface translations, which move the surface along the normals create non-zero flux. Including them results in the extended corner symmetry algebra
\be\label{gext}
\mathfrak{g}^{\rm ext}_S = \left(  \mathrm{diff}(S)\oright\mathfrak{sl(2,\mathbb{R})}^S \right) \oright (\R^2)^S \,.
\ee
This extended algebra has recently been shown to be a universal subgroup  component of the bulk diffeomorphism group in the presence of an embedded surface \cite{Ciambelli:2021vnn}.

The expression for the charges can be worked out by first decomposing a given vector field $\xi$ into its tangential component $\xi_\para=\xi_\para^A\pa_A$ and normal component
$\xi_\perp=\xi_\perp^a \pa_a$, where $\sigma^A$ are coordinates on the sphere while $x^a $ denotes normal coordinates. The sphere being located at $x^a=0$.
The surface boosts corresponds to normal  fields that vanish on $S$: $\xi_\perp^b|_S {=}0$, while $\pa_a \xi_\perp^b|_S {\neq} 0$.
The surface translations corresponds to normal vector fields that do not vanish on $S$.

We introduce a $2+2$ decomposition of the metric in a neighbourhood of $S$, with $A,B$ labeling coordinates tangent to $S$ and $a,b$ coordinates in the normal direction. 
In
coordinates $(x^a, \s^A)$ adapted to the $2+2$  decomposition, the metric reads
\bea \label{metric}
\rd s^2&=&h_{ab}\rd x^a\rd x^b+\gamma_{AB}(\rd \s^A-U^A_a\rd x^a)(\rd \s^B-U^B_b\rd x^b),
\eea
where $\gamma_{AB}$ is the induced metric  on $S$ with determinant $\gamma$, $h_{ab}$ a generalized lapse matrix defining the metric on the normal plane
with determinant $h$, and 
$U^A_a$ a generalized shift, that can be also viewed  
as a normal connection. A basis for normal vectors to $S$ is provided by $\hat{\pa}_a = \pa_a + U^A_a \pa_A$, while $h_{ab}= g(\hat\pa_a,\hat\pa_b)$. 
We work with a parametrization where $\xi= \xi^a\pa_a +\xi^A\pa_A$, with $(\xi^a,\xi^A)$ field independent,\footnote{\la{agno} 
There are two alternative parametrizations that can be naturally considered. The first is
$
\xi = \xi^a \hat{\pa}_a +\hat\xi^A \pa_A,
$ 
 which is related to the one we have by a field-dependent redefinition 
 $\hat{\xi}^A = \xi^A - \xi^a U_a^A$. The second is 
$ \xi = \tilde\xi^a \pa_a  +\xi^A \tilde{\pa}_A $  with $\tilde{\pa}_A =\pa_A + A_A{}^a \pa_a$ and $\tilde{\xi}^a = \xi^a -\xi^A A_A^a$, where $A_A^a= \gamma_{AB} U^B_b \tilde{h}^{ba} $
with $\tilde{h}_{ab}= h_{ab} + \gamma_{AB}U^A_aU^B_b.$
One needs to be aware that the field dependent parametrization changes the algebra from the Lie bracket to the Lie algebroid bracket $\lbr\cdot,\cdot\rbr$.
}
and we can write
\begin{subequations}\begin{align}
& \xi_a= 
h_{ab}\xi^b - \g_{AB}U^A_a(\xi^B -U_b^B \xi^b ) \,,\\
&\xi_A = \gamma_{AB}(\xi^B -U_a^B \xi^a ).
\end{align}\end{subequations}
Let us introduce the bivector normal  $N:= \frac12 N_{\mu\nu } \rd x^\mu \wedge \rd x^\nu = n_0\wedge n_1 $. Its component in the coordinate basis are given by
$N^{ab} = n^{[a}_0 n^{b]}_1  = \epsilon^{ab}$,  
 $N^{AB} = \epsilon^{ab} U^A_a U^B_b$, $N^{aB}=\epsilon^{ab}U^B_b$, where $ \epsilon^{ab}$ is the normal Levi--Civita tensor.\footnote{We use the fact that, introducing the  timelike unit normal $n_0^\m=(n_0^a, n_0^a U_a^A)$ and spacelike unit normal $n_1^\m=(n_1^a, n_1^a U_a^A)$ to $S$.
The normal Levi-Civita tensor is a skew-symmetric tensor such that  $\epsilon_{01} =\sqrt{|h|}$, and $\epsilon^{01}=-1/\sqrt{|h|}$.
}
The Komar charge is given by
\begin{align}
Q^{\textsf K}_\xi
&=\f12\int_S \sqrt{\gamma}N^{\m\n}\pa_\n\xi_\m\cr
&=\f12\int_S
\sqrt{\gamma}\left\{\xi^c \left[  \epsilon^{ab} \hat\p_b {h}_{ac} -  U_c^B \g_{BA} (\epsilon^{ab} \hat\p_a U_b^A) -  D_C ({h}_{ca}\epsilon^{ab} U_b^C) \right]\right.\cr
& \qquad\qquad\qquad+ \left. 
\xi^A  \g_{AB}  (\epsilon^{ab} \hat\p_a U_b^B )
+  \pa_b \xi^c  ({h}_{ca}\epsilon^{ab}) \right\}\,,
\end{align}
where $D_A$ is the covariant derivative compatible with $\gamma_{AB}$.

We see that no component $\pa_a \xi^B$ appears in this expression, which  is why we have a semi-direct product structure.
More explicitly,  if one introduces the following sphere coefficients 
\be
Y^A:=\left.\xi^A\right|_{x^a=0}\,,\qquad W_a{}^b=\left. \pa_a\xi^b\right|_{x^a=0} \,,\qquad T^a= \left.\xi^a\right|_{x^a=0}\,,
\ee
it can be checked that they satisfy the following commutation relations
\begin{align}
[Y,Y']^A&=Y^B \p_BY^{'A}-Y^{'B}\p_B Y^A\,, &[Y,W]_a{}^b&=Y^A\p_A W_a{}^b\,, &[Y,T]^a&=Y^A\p_A T^a\,,\cr
[W,W']_a{}^b&=W_a{}^c W'_c{}^b-W'_a{}^c W_c{}^b\,,\quad &[W,T]^a&=-T^b W_b{}^a\,,\quad   &[T,T']^a&=0\,,
\end{align}
which reproduce the Lie algebra structure \eqref{gext}.
The Komar charge is given by $Q^{\textsf K}_\xi= Q_{(T,W,Y)}$, where
\be \label{KomarTWY}
Q_{(T,W,Y)} =  \int_S \left(Y^A \tilde{P}_A +  W_a{}^b \tilde{N}_b{}^a +  T^a \tilde{Q}_a  \right)\,.
\ee
In the expression above, $\tilde{P}_A= \sqrt{\gamma} P_A$ is  the sphere's momentum density with 
\bea
P_A &=& \f12  \gamma_{AB}  \epsilon^{ab}\hat{\pa}_a U_b^B 
= \f1{2\sqrt{|h|}}\gamma_{AB} (\partial_1U^B_0-\partial_0U^B_1+[U_1,U_0]^B),
\eea
 which is proportional to the curvature of the normal connection. It gives  the generators of the $\mathrm{diff}(S)$ algebra and  
 this is precisely the twist, or non-integrability of the time-like screens.
 The second term is given by the bivector normal density $ \tilde{N}_b{}^a=  \sqrt{\gamma} {N}_b{}^a$ with
 \be
 {N}_b{}^a= \f12  h_{bc}\epsilon^{ca},
\ee
and it yields the  $\mathfrak{sl(2,\mathbb{R})}^S$ algebra generators. 
Finally, the generators of normal translations are given by
$\tilde Q_a = \sqrt{\gamma} Q_a$ with
\begin{align}\la{Qa}
Q_a &=
  \f12 \epsilon^{cb} \hat\p_b {h}_{ac} - U_a^B P_B -  D_C ({N}_a{}^b U_b^C) \,,
\end{align}
and they yield the new $(\R^2)^S$ algebra component corresponding to translations along the 2 normal directions.

The algebra \eqref{gext} for a corner of a finite region of spacetime has been  very recently derived in \cite{Ciambelli:2021vnn}.
Their construction, based on a parametrization of the metric in a neighbourhood of $S$ different from \eqref{metric},  accounts for the  normal translations in the charge bracket by extending the phase space including the embedding information. This construction can be understood in terms of edge modes, in line with the strategy laid out in  \cite{DonnellyFreidel} for symmetry  transformations of the corner, with the embedding map playing the role of the edge modes.
The embedding map allows to reabsorb the field translation into the transformation of the embedding field, which
insures that no flux appears in the  construction of the Poisson bracket; in this way,  the algebra could be canonically represented.  

In our work, the bracket is instead constructed 
using the covariant phase space formalism and we need to work with the generalized Barnich--Troessaert bracket in order to  take into account the presence of symplectic flux and anomalies. 
This provides a  representation of the algebra \eqref{gext} for a finite region, and the same method provides a representation of the BMSW algebra at null infinity \cite{Freidel:2021yqe}.
We notice also that the expression for the charges obtained in \cite{Ciambelli:2021vnn} differs from ours. In particular,
the normal translation charges \eqref{Qa} of the extended corner symmetry coincide with those of \cite{Ciambelli:2021vnn} only in the case $U^A_a=0$. In our context, this choice amounts to a partial gauge-fixing of some metric components at the location of the surface, and preservation of this gauge-fixing would require using a field-dependent parametrization of the diffeomorphisms, as described in footnote \ref{agno}. In the case where  the embedding map is part of the phase space, the step of gauge-fixing is replaced by a choice of special embedding map.  However, we believe that a deeper analysis of the extended phase space including the embedding map and the role of gauge-fixing is still needed.

\subsection{Null infinity}

The set of results presented so far unravels in a clear and precise way the deep connection between Einstein's equations for gravity and the  algebra  of symmetry transformations preserving some given boundary geometrical data. An explicit application of this holographic derivation of Einsteins equations in the case of asymptotic null infinity has been worked out in detail in \cite{Freidel:2021yqe}. This required an extension of the  BMS group \cite{Bondi:1960jsa, BMS, Sachs62}, named \bmw group, that includes also local Weyl rescalings and
arbitrary diffeomorphisms of the 2d sphere metric,  in addition to super-translations, and encompasses previous extensions \cite{Barnich:2009se, Barnich:2011mi, Campiglia:2014yka, Compere:2018ylh}. 
At the same time, the general framework developed here  puts the analysis of finite distance and asymptotic charges and their algebra on equal footing;
this uniformity of treatment then
provides an efficient and clear method for investigating the nature of the celestial sphere symmetry group and possibly reveal its most extended structure. 
In this perspective, it is illustrative to understand how the 
\textsf{bmsw} algebra of residual diffeomorphisms revealed in \cite{Freidel:2021yqe},\footnote{Here $W, T$ correspond to arbitrary functions on the celestial sphere labelling respectively Weyl rescalings (boosts normal to the sphere) and super-translations.}
\be\la{bmsw}
\textsf{bmsw}
= (\mathrm{diff}(S) \oright \R_W^S) \oright \R_T^S\,,
\ee
relates to the extended corner symmetry algebra \eqref{gext}: the term $\R_W^S$ corresponds to the $\sl(2, \R)$ algebra generator that  preserves the null generator of $\scri$, while $\R_T^S$ is one of the two  normal time (super-)translation  generators, namely the one along $\scri$. Therefore, the \textsf{bmsw} Lie algebra represents a subalgebra of the  maximal closed subalgebra $g^{\rm ext}_S$ of the full bulk diffeomorphisms. 

This can be seen explicitly as follows. The Bondi metric can be recast as in \eqref{metric} and it reads as
\be
ds^2 = -2e^{2\beta} \dd u \left(F\dd u + \dd r \right) + r^2 q_{AB} (\rd \s^A-U^A\rd u)(\rd \s^B-U^B\rd u).
\ee
To be more explicit, upon comparison with \eqref{metric}, one has $x^0=u$, $x^1=r$ and
\begin{align}
h_{00} &=-2e^{2\beta}F,\quad &h_{01}&= -e^{2\beta}, \quad &h_{11}&= 0 &\g_{AB} &= r^2 q_{AB}, \quad & U^A_a\dd x^a = U^A\dd u
\end{align}
and $\sqrt{\g}=r^2 \sqrt{q}$, $\sqrt{|h|} = e^{2\beta}$. Thus, the asymptotic expressions for the Komar charge aspects \eqref{KomarTWY} are given as follows.  The generators of the normal translations  give (see Appendix \ref{sec:ST})
\be
\tilde{Q}_0 = \sqrt{\bq}\left( M + \f12 \bD_C \bU^C -\dot{\bar\beta}\right)+o(1), \quad
\tilde{Q}_1 = -\f1 r\sqrt{\bq}~ {\bar\beta} +o(r^{-1})\,.
\ee
We see that they admit a finite limit when $r\to \infty$ and that $\tilde Q_1$ vanishes in this limit.
This is consistent with the fact that the corner symmetry group restricts in this limit to the 
BMSW group investigated in \cite{Freidel:2021yqe}.
The generators of the $\mathrm{diff}(S)$ algebra read
\be
\tilde{P}_A =  \sqrt{\bq}\left[-r\bU_A  + \left(\bP_A + \pa_A \bar\beta\right)\right] +o(1).
\ee
They exhibit a divergent contribution. As shown in \cite{Freidel:2021yqe} the renormalization procedure simply selects the finite part of the momenta as the renormalised symmetry generator.
Finally, the generators of $\mathfrak{sl(2,\mathbb{R})}^S$ are given by 
\be
\tilde{N}_0^{\;0}= -\f{r^2}{2} \sqrt{\bq},\quad
\tilde{N}_0^{\;1}= \sqrt{\bq}\left(r^2\bF - r M\right) + o(r), \quad
\tilde{N}_1^{\;0}= 0,\quad
\tilde{N}_1^{\;1}= -\tilde{N}_0^{\;0}\,.
\ee
These generators diverge and need to be renormalized. 
From this analysis, one expects that after renormalization only one of the $\mathfrak{sl(2,\mathbb{R})}^S$ charges and one the super-translation charges are left non-vanishing,\footnote{Notice in fact that the $\mathfrak{sl(2,\mathbb{R})}^S$ charge $\tilde{N}_0^{\;0}$ has no finite contribution.} reducing the extended corner symmetry algebra to the \textsf{bmsw} one \cite{Freidel:2021yqe}. This expectation is related to the fact that, while  the entanglement sphere at finite distance has two null normals,  there is a preferred null super-translation generator ruling null infinity which plays a central role. 
Let us emphasize that this result suggests  that the extended corner symmetry group can be  bigger  than the asymptotic symmetry group, contrary to the expectation based e.g. on the notion of asymptotic isometries, which have no finite analogue in general spacetimes. 
It would be interesting to perform a more in depth analysis to see if one can recover, in some gravity theories, more than the BMS-like subgroups and have access to some other $\mathfrak{sl(2,\mathbb{R})}^S$ and super-translation components (see also footnote \ref{1}).

We warn the reader  that  some care is needed in comparing this limit with the BMS charges, because those require a specific field dependence of the asymptotic Killing vectors, which was not included here, see discussion in footnote~\ref{agno}.

It is  interesting to notice that in
\cite{Chandrasekaran:2018aop} it was shown that the semi-direct sum structure \eqref{gext} arises also as 
the algebra of symmetries  at a  general non-stationary null surface at finite distance. In this case, the symmetry group preserves a thermal Carrollian geometric structure on the null surface---defined by the equivalence class associated with the null surface generator and the non-affinity smooth function--- with the two $\R$ terms in \eqref{gext} representing  angle-dependent displacements of affine parameter and
as angle-dependent rescalings of affine parameter (see also \cite{Hopfmuller:2016scf, Ciambelli:2019lap, Wieland:2020gno, Chandrasekaran:2020wwn }). This universality features of the symmetry group of a null boundary, being this at finite or infinite distance, is  remarkable and, at the same time, not surprising from the local holography point of view developed here. In fact, this framework can be applied to define a renormalization procedure in a systematic manner, as in \cite{Freidel:2021yqe}, putting the study of finite and infinite surfaces on equal footing. In this way, $\scri$ in the Bondi gauge can be understood as a particular null surface with vanishing non-affinity.

\section{Conclusions}\label{sec:concl}

In this work we have extended the study of the corner symmetry algebra associated with local subsystems of space initiated in \cite{DonnellyFreidel, Freidel:2020xyx, Freidel:2020svx, Freidel:2020ayo}. We have included the {normal} super-translations that move the location of
the corner 2-sphere.
Including this extension to the corner symmetry group is crucial to study the time evolution of subregions.
Time evolution is usually encoded into a boundary and a choice of boundary condition, here we have freed ourselves from these restrictions and looked at the evolution of the corner charges along any normal deformation.
This means that we have an open Hamiltonian system where the symmetry charges are in general non-integrable and we cannot rely on a canonical Poisson bracket to study the extended corner symmetry algebra because of the presence of non-trivial symplectic and Hamiltonian fluxes.

In this paper, we have tackled these issues by  introducing a Noetherian split for the charges and fluxes. 
Our analysis is done for the general case where we allow for Lagrangian and symplectic anomalies \emph{and} an arbitrary dependence of the diffeomorphism generators on the phase space fields. Inclusion of anomalies \cite{Hopfmuller:2018fni} is essential when working with boundaries and the anomaly operator \eqref{ano} represents a key technical tool to develop a consistent covariant phase space formalism. Our analysis extends the work done in \cite{Chandrasekaran:2020wwn} to the case  where the Lagrangian class is non-covariant and the case where the vector fields are field dependent.

One of the key aspect of our construction is the 
definition of a Lagrangian dependent charge bracket which extends the Barnich-Troessaert bracket \cite{Barnich:2010eb} with  a cocycle contribution constructed from the Lagrangian and its  anomaly.
This bracket  presents three striking features. 
First, it can be proven that it satisfies  Jacobi's identity. 
Second, it provides a canonical faithful representation \eqref{Flux1}  of the field-dependent symmetry transformations commutator \eqref{maincom}, avoiding the appearance of cocycles, for any choice of boundary Lagrangian. 
Third, when evaluated off-shell, the identity \eqref{Braoff} shows how the canonical representation of the extended corner symmetry algebra is clearly related to the projection on the sphere of the bulk equations of motion.  
This remarkable connection could be understood as a {\it definition} of local holography. 

We have also shown that the Lagrangian-dependent charge bracket coincides with the canonical bracket.
This follows from the realization that a choice of  Lagrangian within a given class determines a choice of boundary condition and hence that the canonical bracket is also dependent on $L$.

We have then exploited the notion of Noether charge to reveal the fully extended structure of the corner symmetry algebra \eqref{gext}, which includes an additional contribution associated to the translation transformations along the two  directions normal to the corner. 

In spite of these remarkable properties, key open questions remain about this bracket, and should be the goal of future work. First and foremost the central question is whether this bracket can be used as a guide for quantization and realized as a commutator.
One strategy to do this is to  relate the Lagrangian-dependent charge bracket with a canonical bracket for an extended phase space. This means to establish an analog of \eqref{canQ} for a more relaxed set of boundary conditions.
Relaxing the boundary conditions requires introducing boundary edge modes as described in \cite{Freidel:2020svx} in order to extend the phase space.
Preliminary investigations of this strategy of using edge modes to absorb the flux part of our bracket into a Dirac bracket
have appeared in \cite{Ciambelli:2021vnn} using the sphere embedding maps and in \cite{Wieland:2021eth} using null data. Understanding the relations between these different approaches is of crucial importance for the quantization program.

\section*{Acknowledgements}

L. F. would like to thank Luca Ciambelli and Rob Leigh for sharing an early version of their recent results and A. Speranza for discussions on this subject.
Research at Perimeter Institute is supported in part by the Government of Canada through the Department of Innovation, Science and Economic Development Canada and by the Province of Ontario through the Ministry of Colleges and Universities. This project has received funding from the European Union's Horizon 2020 research and innovation programme under the Marie Sklodowska-Curie grant agreement No 841923. 
R.~O.~is  funded  by  the  European  Structural  and  Investment  Funds (ESIF) and the Czech Ministry of Education, Youth and Sports (MSMT), Project CoGraDS-CZ.02.1.01/0.0/0.0/15003/0000437.

\appendix \label{App}
\section{Proofs of the canonical formulae}\la{AppD}
In this appendix we provide the proof of the evaluation \eqref{charge0}, \eqref{Fluxdef}  for  the Lagrangian and symplectic anomalies, 
the proof of the commutation relations \eqref{anomaly},
the proof of the shift property \eqref{Q'} of the charge and flux, and the proof of the main formula \eqref{Braoff} for the charge bracket.
We do not assume here that $E$ or $C_\xi=0$ and keep track of all the bulk contributions.
 
\subsection{Lagrangian and symplectic anomalies}\label{Noetherapp}
One first evaluates the Lagrangian anomaly 
\bea\label{Noetherid}
\Delta_\xi L = \delta_\xi L - \cL_\xi L=  \rd I_\xi \theta - I_\xi E - \rd \iota_\xi L
= \rd (I_\xi \theta -\iota_\xi L - C_\xi)\,.
\eea
The condition $\Delta_\xi L=\rd a_\xi$ implies that the Noether current is 
\be
 I_\xi \theta -\iota_\xi L -a_\xi =  C_\xi + \rd q_\xi.
\ee
One then evaluates the symplectic anomaly:
\bea
\Delta_\xi \theta&=&
(\delta_\xi -{\cal L}_\xi- I_{\delta \xi})\theta \cr
&=& I_\xi \delta \theta + \delta I_\xi\theta   -\iota_\xi \rd \theta - \rd \iota_\xi\theta - I_{\delta \xi}\theta \cr
&=& I_\xi \delta \theta + \delta (I_\xi\theta- \iota_\xi L)  - (I_{\delta \xi}\theta- \iota_{\d\xi} L)
 - \rd \iota_\xi\theta -\iota_\xi E \cr
& {=}& I_\xi \delta \theta + \delta (C_\xi + \rd q_\xi + a_\xi ) -(C_{\delta \xi} + \rd q_{\delta \xi} + a_{\delta \xi} )  - \rd \iota_\xi\theta -\iota_\xi E\,,
\eea
where in the third line we used that $ \iota_\xi \rd \theta =\iota_\xi \delta L +\iota_\xi E=\delta \iota_\xi L - \iota_{\delta\xi} L +\iota_\xi E.$
Using the definition \eqref{covarianceprop} of the symplectic anomaly means that 
\bea
\rd A_\xi = 
I_\xi \delta \theta + \delta (C_\xi + \rd q_\xi  ) -(C_{\delta \xi} + \rd q_{\delta \xi} )  - \rd \iota_\xi\theta -\iota_\xi E.
\eea
After integration over $\Sigma$ this means that 
\be 
-I_\xi \Omega = \int_{\Sigma} (\delta C_\xi - C_{\delta \xi}-\iota_\xi E)
+\int_S (\delta q_\xi - q_{\delta \xi} - \iota_\xi\theta - A_\xi).
\ee
This equation can be written as the sum of an integrable piece and a flux component
\be 
-I_\xi \Omega = \delta \left( \int_{\Sigma}  C_\xi +\int_S  q_\xi \right) 
- \int_{\Sigma}( C_{\delta \xi}+ \iota_\xi E)
-\int_S ( q_{\delta \xi} + \iota_\xi\theta + A_\xi)\,.
\ee
The total charge is a sum of a constraint and a boundary charge and similarly
the flux is also the sum of a bulk contribution and a boundary contribution.
It is interesting to note that the bulk contribution of the flux vanishes if $\xi$ is field independent \emph{and} is tangent to $\Sigma$.
These are the kinematical symmetries.
It was already noted in \cite{Freidel:2020svx} that the kinematical symmetries plays a special  role in the canonical analysis.

\subsection{Lagrangian shift}\la{App:Lshift}

Here we provide the proof of the expressions \eqref{Q'}  in the case where the  Lagrangian is modified by a boundary term such that
\be
L'-L=\rd \ell,\qquad \theta' -\theta =   \delta \ell -\rd \vartheta\,,
\ee
and
\be
\Delta_\xi \theta'=\Delta_\xi \theta+\Delta_\xi \delta \ell -\Delta_\xi\rd \vartheta\,.
\ee
We have 
\be
\delta L'= \delta L + \rd \delta \ell= \rd \theta + \rd (  \theta' -\theta) + E_L
= \rd \theta' + E_L\,,
\ee
and, on-shell of the equations of motion, 
we have
\bea
\Delta_\xi \theta'
&=& I_\xi \Omega' + \delta (I_\xi\theta'- \iota_\xi L') - \rd \iota_\xi\theta'+\iota_{\d\xi} L' - I_{\delta \xi}\theta' \,.
\eea
Therefore, the balance formula
\be
-I_\xi \Omega' =  \delta Q'_{\xi}-\cF'_{\xi} 
\ee
yields the modified charge
\bea
\rd q'_\xi&=&I_\xi\theta'- \iota_\xi L'-a_\xi-\Delta_\xi  \ell \cr
&=&\rd q_\xi +I_\xi\delta \ell -I_\xi\rd \vartheta 
- \iota_\xi \rd\ell
-\Delta_\xi  \ell\cr
&=&\rd (q_\xi+\iota_\xi \ell  -I_\xi \vartheta )\,,
\eea
which implies that 
\be
q'_\xi- q_\xi = \iota_\xi \ell  -I_\xi \vartheta.
\ee
The difference in flux is given 
\bea
\cF'_\xi -\cF_\xi&=& \int_S \left(\iota_\xi (\theta' -\theta) + q'_{\delta \xi}- q_{\delta\xi} -  \Delta_\xi \vartheta\right)\cr
&=&  \int_S\left ( \iota_\xi\delta \ell -\iota_\xi\rd \vartheta
+\iota_{\delta \xi} \ell  -I_{\delta \xi} \vartheta -  \Delta_\xi \vartheta\right)\cr
&=&\int_S\left (
   \d  \iota_\xi \ell-(\Delta_\xi +\cL_\xi+I_{\delta \xi})\vartheta
\right)\cr
&=&\int_S\left ( \d  \iota_\xi \ell-\d_\xi \vartheta
\right)\,.
\eea

\subsection{Cartan commutators}\la{CartanC}

Let us  also derive the commutation relations \eqref{anomaly}. By means of \eqref{Com1}, \eqref{vbra},
 and
\begin{align}
&[I_{\delta \xi},  I_{\chi}]=- I_{\delta_\chi \xi}\,,\quad [\cL_\xi, I_\chi]=0\,,\quad[\d_\xi, I_{\chi}]=-I_{\lbr\xi, \chi\rbr}\,,\cr
& [I_{\delta \xi},  \iota_{\chi}]=0\,,\quad [\cL_\xi, \iota_\chi]= \iota_{[\xi,\chi]_{\mathrm{Lie}}}\,,\quad [\d_\xi, \iota_{\chi}]=\iota_{\d_\xi \chi}\,,
\end{align}
we have
\bea
[\Delta_\xi, I_{\chi}]&=&[\delta_\xi,I_{\chi} ]  - [I_{\delta \xi},I_{\chi} ]
=I_{\delta_\chi\xi} -I_{\lbr\xi,\chi\rbr}\,,\cr
[\Delta_\xi, \iota_{\chi}]&=&[\delta_\xi,\iota_{\chi} ]-[\cL_\xi, \iota_\chi]
=\iota_{\d_\xi \chi}-\iota_{[\xi,\chi]_{\mathrm{Lie}}}
=\iota_{\delta_\chi\xi} -\iota_{\lbr\xi,\chi\rbr}\,,
\eea
The two commutators above also yield
\bea
[\Delta_\xi , \d_\chi]&=&\d [\Delta_\xi ,I_\chi]+ [\Delta_\xi ,I_\chi]\d
+[\Delta_\xi ,\d]I_\chi+I_\chi[\Delta_\xi ,\d]
\\
&=& \d_{\delta_\chi\xi} -\d_{\lbr\xi,\chi\rbr}-\Delta_{\d_\chi \xi}\cr
&=& \cL_{\delta_\chi\xi} -\d_{\lbr\xi,\chi\rbr}\,,
\cr
[\Delta_\xi , \cL_\chi]&=&\rd [\Delta_\xi ,\iota_\chi]+ [\Delta_\xi ,\iota_\chi]\rd
= \cL_{\delta_\chi\xi} - \cL_{\lbr\xi,\chi\rbr}\,,
\eea
from which follows
\be
[\Delta_\xi ,\Delta_\chi]= -\Delta_{\lbr\xi,\chi\rbr}\,.
\ee

\subsection{Main formula}
\label{Mainf}
Here we derive \eqref{Braoff}. We first  provide the proof of \eqref{QAnomaly}.
One starts by the evaluation of
\bea
\Delta_\xi I_\chi \theta&=& [\Delta_\xi, I_\chi] \theta + I_\chi  \Delta_\xi \theta\cr
&=& I_{\delta_\chi\xi}\theta  -I_{\lbr\xi,\chi\rbr}\theta + \delta_\chi a_\xi -a_{\delta_\chi \xi} + \rd I_\chi A_\xi.
\eea
Similarly 
\bea
\Delta_\xi \iota_\chi L&=& [\Delta_\xi, \iota_\chi] L + \iota_\chi  \Delta_\xi L \cr
&=& \iota_{\delta_\chi\xi}L -\iota_{\lbr\xi,\chi\rbr}L+ \cL_\chi a_\xi -   \rd \iota_\chi a_\xi
\eea
Therefore, taking the difference and using \eqref{Noetherapp} we obtain that 
\bea
\Delta_\xi( C_\chi + a_\chi +\rd q_\chi)&=&
( C_{\delta_\chi\xi}  +\rd q_{\delta_\chi\xi})
- 
( C_{\lbr\xi,\chi\rbr} + a_{\lbr\xi,\chi\rbr} +\rd q_{\lbr\xi,\chi\rbr})\cr
&+& \Delta_{\chi} a_\xi + \rd (I_\chi A_\xi +\iota_\chi a_\xi).
\eea
which can be recast as
\bea
\Delta_\xi( C_\chi +\rd q_\chi)&=&
( C_{\delta_\chi\xi}  +\rd q_{\delta_\chi\xi})
- 
( C_{\lbr\xi,\chi\rbr}  +\rd q_{\lbr\xi,\chi\rbr})\cr
&-& ( \Delta_\xi a_{\chi} -\Delta_{\chi} a_\xi +a_{\lbr\xi,\chi\rbr})+ \rd (I_\chi A_\xi +\iota_\chi a_\xi).
\eea
From which we conclude using the definition of $c_{(\xi,\chi)}$  that 
\bea
\Delta_\xi( C_\chi +\rd q_\chi) -
( C_{\delta_\chi\xi}  +\rd q_{\delta_\chi\xi})&=&
- 
( C_{\lbr\xi,\chi\rbr}  +\rd q_{\lbr\xi,\chi\rbr})+ \rd (I_\chi A_\xi +\iota_\chi a_\xi- c_{(\xi,\chi)}).
\eea

One then uses that $\tilde\cF_\xi=\int_{\Sigma}  \tilde{f}_\xi$ with 
\bea
\tilde{f}_\xi&=& C_{\delta \xi}+\iota_\xi E
+\rd  ( q_{\delta \xi} + \iota_\xi\theta + A_\xi).
\eea
Therefore 
\bea
I_\chi \tilde{f}_\xi -\rd \iota_\xi\iota_\chi L&=&
C_{\delta_\chi \xi}+\iota_\xi I_\chi E
+\rd  ( q_{\delta_\chi \xi} + \iota_\xi(I_\chi \theta-\iota_\chi L)  + I_\chi A_\xi)\cr
&=& ( C_{\delta_\chi\xi}  +\rd q_{\delta_\chi\xi}) +
\rd \iota_\xi( C_\chi +\rd q_{\chi} +a_\chi)
+\rd (I_\chi A_\xi) +\iota_\xi \rd C_\chi \cr
&=& ( C_{\delta_\chi\xi}  +\rd q_{\delta_\chi\xi}) +
\cL_\xi( C_\chi +\rd q_{\chi} )
+\rd (\iota_\xi a_\chi+ I_\chi A_\xi) \cr
&=& \delta_\xi( C_\chi +\rd q_{\chi} )
+ ( C_{\delta_\chi\xi}  +\rd q_{\delta_\chi\xi}) -
\Delta_\xi( C_\chi +\rd q_{\chi} )
+\rd (\iota_\xi a_\chi+ I_\chi A_\xi) \cr
&=& \delta_\xi( C_\chi +\rd q_{\chi} )
+ ( C_{\lbr\xi,\chi\rbr}  +\rd q_{\lbr\xi,\chi\rbr})
+ \rd (\iota_\xi a_\chi -\iota_\chi a_\xi+ c_{(\xi,\chi)}),
\eea
which gives us 
\bea
\delta_\xi( C_\chi +\rd q_{\chi} ) -
I_\chi \tilde{f}_\xi +\rd (\iota_\xi\iota_\chi L + \iota_\xi a_\chi -\iota_\chi a_\xi+ c_{(\xi,\chi)} )
= - C_{\lbr\xi,\chi\rbr}  -\rd q_{\lbr\xi,\chi\rbr}.
\eea
We can split this contribution into a bulk contribution depending on $C_\xi$ and a boundary contribution.
The bulk contribution of the LHS reads
\bea
 \delta_\xi C_\chi - C_{\delta_\chi  \xi}- \iota_\xi \rd C_\chi 
 = \Delta_\xi C_\chi -
 C_{\delta_\chi  \xi}+  \rd \iota_\xi  C_\chi . 
 \eea

The boundary flux 2-form  is 
\be
f_\xi :=  q_{\delta \xi} + \iota_\xi\theta + A_\xi
\ee
and the differential form of the bracket is 
\be
\{q_\xi,q_{\chi}\} :=  \delta_\xi q_\chi - I_\chi f_\xi  + (\iota_\xi\iota_\chi L +\iota_\xi a_\chi -\iota_\chi a_\xi+ c_{(\xi,\chi)} ).
\ee
The previous equation therefore reads
\be
  \rd\left( \iota_\xi  C_\chi  + 
 \{q_\xi,q_{\chi}\} +  q_{\lbr\xi,\chi\rbr}\right) = -\left( \Delta_\xi C_\chi -
 C_{\delta_\chi  \xi} + C_{\lbr\xi,\chi\rbr}  \right).
\ee
The RHS of this equation contains the anomaly of the constraint. By means of the relation \eqref{DC}, we thus recover the main formula \eqref{Braoff}.

\subsection{Relation bracket-symplectic form}\label{BracketS}

Here we  provide the derivation of the relation \eqref{Oxc}. 
For this  proof we go back on-shell and use $E\heq0$.
As we are going to use the relation
\be\la{ddL}
\d_\chi \iota_\xi L - \d_\xi \iota_\chi L + \iota_{\lbr\chi, \xi\rbr} L=  \iota_\xi \rd a_\chi - \iota_\chi \rd a_\xi
 -\rd \iota_\xi \iota_\chi L\,,
\ee
let us first prove this.
 By means of
\be
[\cL_\xi, \iota_\chi]=\iota_{[\xi,\chi]}\,,\quad [\d_\xi, \iota_{\chi}]=\iota_{\d_\xi \chi}\,,\quad 
  \lbr\xi,\chi\rbr:= [\xi, \chi]_{\mathrm{Lie}} + \delta_\chi \xi - \delta_\xi \chi\,,
\ee
we have
\bea
\d_\chi \iota_\xi  - \d_\xi \iota_\chi  + \iota_{\lbr\chi, \xi\rbr} &=&
\iota_\xi\d_\chi   + \iota_{\d_\chi \xi } -  \iota_\chi \d_\xi - \iota_{\d_\xi \chi } + \iota_{\lbr\chi, \xi\rbr} \cr
&=&\iota_\xi\cL_\chi +\iota_\xi\Delta _\chi +\iota_\xi I_{\d_\chi} -  \iota_\chi \cL_\xi -  \iota_\chi \Delta_\xi -\iota_\chi I_{\d_\xi}+ \iota_{[\chi, \xi]} 
+\iota_\xi I_{\d_\chi}-\iota_\chi I_{\d_\xi}\cr
&=&\cL_\chi  \iota_\xi-\iota_{[\chi,\xi]} -  \iota_\chi \cL_\xi +\iota_\xi\Delta _\chi  -  \iota_\chi \Delta_\xi + \iota_{[\chi, \xi]}
+\iota_\xi I_{\d_\chi}-\iota_\chi I_{\d_\xi}\ \cr
&=&\rd \iota_\chi  \iota_\xi +\iota_\chi \rd \iota_\xi-  \iota_\chi \rd \iota _\xi -\iota_\chi  \iota _\xi\rd+\iota_\xi\Delta _\chi  -  \iota_\chi \Delta_\xi
+\iota_\xi I_{\d_\chi}-\iota_\chi I_{\d_\xi}\cr
&=&\rd \iota_\chi  \iota_\xi -\iota_\chi  \iota _\xi\rd+\iota_\xi\Delta _\chi  -  \iota_\chi \Delta_\xi
+\iota_\xi I_{\d_\chi}-\iota_\chi I_{\d_\xi}\,,
\eea
from which
\bea
\d_\chi \iota_\xi L - \d_\xi \iota_\chi L + \iota_{\lbr\chi, \xi\rbr} L&=&
\rd \iota_\chi  \iota_\xi L -\iota_\chi  \iota _\xi\rd L+\iota_\xi\Delta _\chi L -  \iota_\chi \Delta_\xi L\cr
&=& \iota_\xi\rd a _\chi  -  \iota_\chi \rd a_\xi-\rd \iota_\xi  \iota_\chi L\,. 
\eea

We can now use the commutator \eqref{DI} and the relation \eqref{ddL} to compute
\bea
I_\xi I_\chi \Omega&\heq& \int_\Sigma I_\xi \d_\chi \theta -\int_\Sigma I_\xi \d(\rd q_\chi +\iota_\chi L +a_\chi)\cr
&=& \int_\Sigma\d_\chi   I_\xi  \theta + \int_\Sigma I_{\lbr\chi, \xi\rbr} \theta
-\int_S \d_\xi q_\chi  -\int_\Sigma \d_\xi \iota_\chi L  -\int_\Sigma \d_\xi a_\chi\cr
&\heq&  \int_S \d_\chi  q_\xi +\int_\Sigma \d_\chi \iota_\xi L +\int_\Sigma \d_\chi  a_\xi\cr
 &+&\int_S  q_{\lbr\chi, \xi\rbr} +\int_\Sigma \iota_{\lbr\chi, \xi\rbr}L +\int_\Sigma   a_{\lbr\chi, \xi\rbr}\cr
&-&\int_S  \d_\xi q_\chi  -\int_\Sigma \d_\xi \iota_\chi L  -\int_\Sigma \d_\xi a_\chi\cr
&=&  \int_S  \left( \d_\chi  q_\xi-\d_\xi q_\chi +q_{\lbr\chi, \xi\rbr} +\iota_\chi \iota_\xi  L\right)\cr
&+&\int_\Sigma (  \iota_\xi \rd a_\chi- \iota_\chi \rd a_\xi )
+\int_\Sigma ( \cL_\chi a_\xi -\cL_\xi a_\chi )
+\int_\Sigma (\Delta_\chi  a_\xi - \Delta_\xi a_\chi + a_{\lbr\chi, \xi\rbr})\cr
&=&   \int_S  \left( \d_\chi  q_\xi-\d_\xi q_\chi +q_{\lbr\chi, \xi\rbr} +\iota_\chi \iota_\xi  L\right)\cr
&+&   \int_S  \left( \iota_\chi a_\xi -\iota_\xi a_\chi  - c_{(\xi,\chi)}\right)\,,
\eea
which thus proves \eqref{Oxc}.

\subsection{Super-translation charges}\la{sec:ST}

We use the expression for the time translation charges 
\be
\tilde Q_a =\sqrt{\gamma}
\left(
 \f12 \epsilon^{cb} \hat\p_b {h}_{ac} - U_a^B P_B -   D_C ({N}_a{}^b U_b^C)\right)\,,
\ee
with
\be
{N}_b{}^a= \f12 h_{bc}N^{ca}\,,
\ee
and  (using the convention of \cite{Freidel:2021yqe} for the metric in Bondi--Sachs coordinates)
\begin{align}
h_{00} &=-2e^{2\beta}F,\quad h_{01}= -e^{2\beta}, \quad h_{11}= 0\,,\quad  \g_{AB} = r^2 q_{AB}\,, \cr  
U^A_a\dd x^a &= U_0^A\dd u\,,
\quad \epsilon^{01}=-e^{-2\beta}\,,\quad 
\sqrt{\g}=r^2 \sqrt{q}\,,\quad \sqrt{|h|} = e^{2\beta}\,,\cr
U_0^A&=\frac{\bar{U}^A}{r^2}- 
\frac{2}{ 3 r^3} \bar{q}^{AB}\left(\bP_B+ \bC_{BC}\bU^C +\pa_B\bar\beta\right) +o(r^{-3})\,,\quad U^A_1=0\,,\cr
\beta&=\frac{\bar\beta}{r^2}+o(r^{-2})\,,\quad F= \bar{F}- \frac{  M}{r}+o(r^{-1})\,,
\end{align}
to compute the two super-translation charges in the limit $r\rightarrow \infty$
\bea
\tilde Q_0& =&\sqrt{\gamma}
\left(
 \f12 \epsilon^{01} \hat\p_1 {h}_{00}+  \f12\epsilon^{10} \hat\p_0 {h}_{01}
- U_0^A P_A -   D_C ({N}_0{}^0 U_0^C)\right)\cr
&=&\f12 \f{\sqrt{\gamma}}{\sqrt{|h|}}
\left(
 \hat\p_0 {h}_{01} -   \hat\p_1 {h}_{00} \right)
 -U_0^A \tilde P_A
 - \f12 \sqrt{\gamma}D_C \left(\f1{\sqrt{|h|}}h_{01}U_0^C\right)\cr
 &=&\f12 \f{\sqrt{\gamma}}{\sqrt{|h|}}
\left(
 \p_0 {h}_{01} +U^A_0\p_A {h}_{01} -   \p_1 {h}_{00} \right)
 +\f12  \sqrt{\gamma}D_AU_0^A
 +o(1)\cr
 &=&\f12 {r^2}\sqrt{q} 
\left(
 - 2\dot \beta -2U^A_0\p_A \beta +2 (\p_r F +2F \p_r\beta)   \right)
 +\f12  \sqrt{q}D_A \bar U^A
 +o(1)\cr
 &=&
\sqrt{q} \left(M - \dot {\bar\beta}+\f 12 D_A \bar U^A\right)+o(1)\,,
\eea
and
\bea
\tilde Q_1&=&
\sqrt{\gamma}
\left(
\f{1}{2}  N^{01} \hat \p_1 {h}_{10} -   D_C ({N}_1{}^0 U_0^C)\right)\cr
&=&
\f{ r^2}{2} \sqrt{q}
 N^{01} \p_1 {h}_{10} \cr
 &=&-\f 1r\sqrt{q} \bar\beta\,.
\eea

\newpage
\bibliographystyle{bib-style2}
\bibliography{biblio-fluxes}

\providecommand{\href}[2]{#2}\begingroup\raggedright\begin{thebibliography}{10}

\bibitem{DonnellyFreidel}
W.~Donnelly and L.~Freidel, \emph{{Local subsystems in gauge theory and
  gravity}}, \href{http://dx.doi.org/10.1007/JHEP09(2016)102}{\emph{JHEP}
  {\bfseries 09} (2016) 102},
  [\href{https://arxiv.org/abs/1601.04744}{{\ttfamily 1601.04744}}].

\bibitem{Freidel:2015gpa}
L.~Freidel and A.~Perez, \emph{{Quantum gravity at the corner}},
  \href{http://dx.doi.org/10.3390/universe4100107}{\emph{Universe} {\bfseries
  4} (2018) 107}, [\href{https://arxiv.org/abs/1507.02573}{{\ttfamily
  1507.02573}}].

\bibitem{Freidel:2016bxd}
L.~Freidel, A.~Perez and D.~Pranzetti, \emph{{Loop gravity string}},
  \href{http://dx.doi.org/10.1103/PhysRevD.95.106002}{\emph{Phys. Rev.}
  {\bfseries D95} (2017) 106002},
  [\href{https://arxiv.org/abs/1611.03668}{{\ttfamily 1611.03668}}].

\bibitem{Freidel:2018pvm}
L.~Freidel and E.~R. Livine, \emph{{Bubble networks: framed discrete geometry
  for quantum gravity}},
  \href{http://dx.doi.org/10.1007/s10714-018-2493-y}{\emph{Gen. Rel. Grav.}
  {\bfseries 51} (2019) 9}, [\href{https://arxiv.org/abs/1810.09364}{{\ttfamily
  1810.09364}}].

\bibitem{Freidel:2019ees}
L.~Freidel, E.~R. Livine and D.~Pranzetti, \emph{{Gravitational edge modes:
  from Kac-Moody charges to Poincar{\'e} networks}},
  \href{http://dx.doi.org/10.1088/1361-6382/ab40fe}{\emph{Class. Quant. Grav.}
  {\bfseries 36} (2019) 195014},
  [\href{https://arxiv.org/abs/1906.07876}{{\ttfamily 1906.07876}}].

\bibitem{Freidel:2019ofr}
L.~Freidel, E.~R. Livine and D.~Pranzetti, \emph{{Kinematical Gravitational
  Charge Algebra}},
  \href{http://dx.doi.org/10.1103/PhysRevD.101.024012}{\emph{Phys. Rev. D}
  {\bfseries 101} (2020) 024012},
  [\href{https://arxiv.org/abs/1910.05642}{{\ttfamily 1910.05642}}].

\bibitem{Freidel:2020xyx}
L.~Freidel, M.~Geiller and D.~Pranzetti, \emph{{Edge modes of gravity - I:
  Corner potentials and charges}},
  [\href{https://arxiv.org/abs/2006.12527}{{\ttfamily 2006.12527}}].

\bibitem{Freidel:2020svx}
L.~Freidel, M.~Geiller and D.~Pranzetti, \emph{{Edge modes of gravity - II:
  Corner metric and Lorentz charges}},
  [\href{https://arxiv.org/abs/2007.03563}{{\ttfamily 2007.03563}}].

\bibitem{Freidel:2020ayo}
L.~Freidel, M.~Geiller and D.~Pranzetti, \emph{{Edge modes of gravity. Part
  III. Corner simplicity constraints}},
  \href{http://dx.doi.org/10.1007/JHEP01(2021)100}{\emph{JHEP} {\bfseries 01}
  (2021) 100}, [\href{https://arxiv.org/abs/2007.12635}{{\ttfamily
  2007.12635}}].

\bibitem{Donnelly:2020xgu}
W.~Donnelly, L.~Freidel, S.~F. Moosavian and A.~J. Speranza,
  \emph{{Gravitational Edge Modes, Coadjoint Orbits, and Hydrodynamics}},
  [\href{https://arxiv.org/abs/2012.10367}{{\ttfamily 2012.10367}}].

\bibitem{Ciambelli:2021vnn}
L.~Ciambelli and R.~G. Leigh, \emph{{Isolated Surfaces and Symmetries of
  Gravity}},  [\href{https://arxiv.org/abs/2104.07643}{{\ttfamily
  2104.07643}}].

\bibitem{Kijowski1976ACS}
J.~D. Kijowski and W.~Szczyrba, \emph{A canonical structure for classical field
  theories}, {\emph{Communications in Mathematical Physics} {\bfseries 46}
  (1976) 183--206}.

\bibitem{Crnkovic:1986ex}
C.~Crnkovic and E.~Witten, \emph{{Covariant description of canonical formalism
  in geometrical theories}}, pp.~676--684.
\newblock Three Hundred Years of Gravitation, Cambridge: Cambridge University
  Press, 1987, pp. 676--684, 1986.

\bibitem{Ashtekar:1990gc}
A.~Ashtekar, L.~Bombelli and O.~Reula, \emph{The covariant phase space of
  asymptotically flat gravitational fields},  in \emph{Mechanics, Analysis and
  Geometry: 200 Years After Lagrange} (M.~Francaviglia, ed.), North-Holland
  Delta Series, pp.~417 -- 450.
\newblock Elsevier, Amsterdam, 1991.
\newblock \href{http://dx.doi.org/10.1016/B978-0-444-88958-4.50021-5}{DOI}.

\bibitem{Lee:1990nz}
J.~Lee and R.~M. Wald, \emph{{Local symmetries and constraints}},
  \href{http://dx.doi.org/10.1063/1.528801}{\emph{J. Math. Phys.} {\bfseries
  31} (1990) 725--743}.

\bibitem{Barnich:1991tc}
G.~Barnich, M.~Henneaux and C.~Schomblond, \emph{{On the covariant description
  of the canonical formalism}},
  \href{http://dx.doi.org/10.1103/PhysRevD.44.R939}{\emph{Phys. Rev. D}
  {\bfseries 44} (1991) R939--R941}.

\bibitem{Wald:1999wa}
R.~M. Wald and A.~Zoupas, \emph{{A General definition of 'conserved quantities'
  in general relativity and other theories of gravity}},
  \href{http://dx.doi.org/10.1103/PhysRevD.61.084027}{\emph{Phys. Rev. D}
  {\bfseries 61} (2000) 084027},
  [\href{https://arxiv.org/abs/gr-qc/9911095}{{\ttfamily gr-qc/9911095}}].

\bibitem{Ashtekar:1981bq}
A.~Ashtekar and M.~Streubel, \emph{{Symplectic Geometry of Radiative Modes and
  Conserved Quantities at Null Infinity}},
  \href{http://dx.doi.org/10.1098/rspa.1981.0109}{\emph{Proc. Roy. Soc. Lond.
  A} {\bfseries 376} (1981) 585--607}.

\bibitem{Dray:1984rfa}
T.~Dray and M.~Streubel, \emph{{Angular momentum at null infinity}},
  \href{http://dx.doi.org/10.1088/0264-9381/1/1/005}{\emph{Class. Quant. Grav.}
  {\bfseries 1} (1984) 15--26}.

\bibitem{Barnich:2001jy}
G.~Barnich and F.~Brandt, \emph{{Covariant theory of asymptotic symmetries,
  conservation laws and central charges}},
  \href{http://dx.doi.org/10.1016/S0550-3213(02)00251-1}{\emph{Nucl. Phys.}
  {\bfseries B633} (2002) 3--82},
  [\href{https://arxiv.org/abs/hep-th/0111246}{{\ttfamily hep-th/0111246}}].

\bibitem{Barnich:2004uw}
G.~Barnich and G.~Compere, \emph{{Generalized Smarr relation for Kerr AdS black
  holes from improved surface integrals}},
  \href{http://dx.doi.org/10.1103/PhysRevD.73.029904}{\emph{Phys. Rev. D}
  {\bfseries 71} (2005) 044016},
  [\href{https://arxiv.org/abs/gr-qc/0412029}{{\ttfamily gr-qc/0412029}}].

\bibitem{Barnich:2007bf}
G.~Barnich and G.~Compere, \emph{{Surface charge algebra in gauge theories and
  thermodynamic integrability}},
  \href{http://dx.doi.org/10.1063/1.2889721}{\emph{J. Math. Phys.} {\bfseries
  49} (2008) 042901}, [\href{https://arxiv.org/abs/0708.2378}{{\ttfamily
  0708.2378}}].

\bibitem{Barnich:2009se}
G.~Barnich and C.~Troessaert, \emph{{Symmetries of asymptotically flat 4
  dimensional spacetimes at null infinity revisited}},
  \href{http://dx.doi.org/10.1103/PhysRevLett.105.111103}{\emph{Phys. Rev.
  Lett.} {\bfseries 105} (2010) 111103},
  [\href{https://arxiv.org/abs/0909.2617}{{\ttfamily 0909.2617}}].

\bibitem{Barnich:2010eb}
G.~Barnich and C.~Troessaert, \emph{{Aspects of the BMS/CFT correspondence}},
  \href{http://dx.doi.org/10.1007/JHEP05(2010)062}{\emph{JHEP} {\bfseries 05}
  (2010) 062}, [\href{https://arxiv.org/abs/1001.1541}{{\ttfamily 1001.1541}}].

\bibitem{Barnich:2011mi}
G.~Barnich and C.~Troessaert, \emph{{BMS charge algebra}},
  \href{http://dx.doi.org/10.1007/JHEP12(2011)105}{\emph{JHEP} {\bfseries 12}
  (2011) 105}, [\href{https://arxiv.org/abs/1106.0213}{{\ttfamily 1106.0213}}].

\bibitem{Barnich:2013axa}
G.~Barnich and C.~Troessaert, \emph{{Comments on holographic current algebras
  and asymptotically flat four dimensional spacetimes at null infinity}},
  \href{http://dx.doi.org/10.1007/JHEP11(2013)003}{\emph{JHEP} {\bfseries 11}
  (2013) 003}, [\href{https://arxiv.org/abs/1309.0794}{{\ttfamily 1309.0794}}].

\bibitem{Compere:2018ylh}
G.~Comp\`{e}re, A.~Fiorucci and R.~Ruzziconi, \emph{{Superboost transitions,
  refraction memory and super-Lorentz charge algebra}},
  \href{http://dx.doi.org/10.1007/JHEP11(2018)200}{\emph{JHEP} {\bfseries 11}
  (2018) 200}, [\href{https://arxiv.org/abs/1810.00377}{{\ttfamily
  1810.00377}}].

\bibitem{Compere:2020lrt}
G.~Comp\`ere, A.~Fiorucci and R.~Ruzziconi, \emph{{The $\Lambda$-BMS$_4$ charge
  algebra}}, \href{http://dx.doi.org/10.1007/JHEP10(2020)205}{\emph{JHEP}
  {\bfseries 10} (2020) 205},
  [\href{https://arxiv.org/abs/2004.10769}{{\ttfamily 2004.10769}}].

\bibitem{Donnay:2015abr}
L.~Donnay, G.~Giribet, H.~A. Gonzalez and M.~Pino, \emph{{Supertranslations and
  Superrotations at the Black Hole Horizon}},
  \href{http://dx.doi.org/10.1103/PhysRevLett.116.091101}{\emph{Phys. Rev.
  Lett.} {\bfseries 116} (2016) 091101},
  [\href{https://arxiv.org/abs/1511.08687}{{\ttfamily 1511.08687}}].

\bibitem{Donnay:2016ejv}
L.~Donnay, G.~Giribet, H.~A. Gonz\'alez and M.~Pino, \emph{{Extended Symmetries
  at the Black Hole Horizon}},
  \href{http://dx.doi.org/10.1007/JHEP09(2016)100}{\emph{JHEP} {\bfseries 09}
  (2016) 100}, [\href{https://arxiv.org/abs/1607.05703}{{\ttfamily
  1607.05703}}].

\bibitem{Donnay:2019jiz}
L.~Donnay and C.~Marteau, \emph{{Carrollian Physics at the Black Hole
  Horizon}}, \href{http://dx.doi.org/10.1088/1361-6382/ab2fd5}{\emph{Class.
  Quant. Grav.} {\bfseries 36} (2019) 165002},
  [\href{https://arxiv.org/abs/1903.09654}{{\ttfamily 1903.09654}}].

\bibitem{Hopfmuller:2016scf}
F.~Hopfmuller and L.~Freidel, \emph{{Gravity Degrees of Freedom on a Null
  Surface}}, \href{http://dx.doi.org/10.1103/PhysRevD.95.104006}{\emph{Phys.
  Rev.} {\bfseries D95} (2017) 104006},
  [\href{https://arxiv.org/abs/1611.03096}{{\ttfamily 1611.03096}}].

\bibitem{Hopfmuller:2018fni}
F.~Hopfm{\"u}ller and L.~Freidel, \emph{{Null Conservation Laws for Gravity}},
  [\href{https://arxiv.org/abs/1802.06135}{{\ttfamily 1802.06135}}].

\bibitem{Adami:2020amw}
H.~Adami, D.~Grumiller, S.~Sadeghian, M.~M. Sheikh-Jabbari and C.~Zwikel,
  \emph{{T-Witts from the horizon}},
  \href{http://dx.doi.org/10.1007/JHEP04(2020)128}{\emph{JHEP} {\bfseries 04}
  (2020) 128}, [\href{https://arxiv.org/abs/2002.08346}{{\ttfamily
  2002.08346}}].

\bibitem{Grumiller:2019ygj}
D.~Grumiller, M.~M. Sheikh-Jabbari, C.~Troessaert and R.~Wutte,
  \emph{{Interpolating Between Asymptotic and Near Horizon Symmetries}},
  \href{http://dx.doi.org/10.1007/JHEP03(2020)035}{\emph{JHEP} {\bfseries 03}
  (2020) 035}, [\href{https://arxiv.org/abs/1911.04503}{{\ttfamily
  1911.04503}}].

\bibitem{Grumiller:2020vvv}
D.~Grumiller, M.~M. Sheikh-Jabbari and C.~Zwikel, \emph{{Horizons 2020}},
  \href{http://dx.doi.org/10.1142/S0218271820430063}{\emph{Int. J. Mod. Phys.
  D} {\bfseries 29} (2020) 2043006},
  [\href{https://arxiv.org/abs/2005.06936}{{\ttfamily 2005.06936}}].

\bibitem{Chandrasekaran:2018aop}
V.~Chandrasekaran, E.~E. Flanagan and K.~Prabhu, \emph{{Symmetries and charges
  of general relativity at null boundaries}},
  \href{http://dx.doi.org/10.1007/JHEP11(2018)125}{\emph{JHEP} {\bfseries 11}
  (2018) 125}, [\href{https://arxiv.org/abs/1807.11499}{{\ttfamily
  1807.11499}}].

\bibitem{Chandrasekaran:2020wwn}
V.~Chandrasekaran and A.~J. Speranza, \emph{{Anomalies in gravitational charge
  algebras of null boundaries and black hole entropy}},
  \href{http://dx.doi.org/10.1007/JHEP01(2021)137}{\emph{JHEP} {\bfseries 01}
  (2021) 137}, [\href{https://arxiv.org/abs/2009.10739}{{\ttfamily
  2009.10739}}].

\bibitem{Compere:2019bua}
G.~Comp\`ere, A.~Fiorucci and R.~Ruzziconi, \emph{{The $\Lambda$-BMS$_4$ group
  of dS$_4$ and new boundary conditions for AdS$_4$}},
  \href{http://dx.doi.org/10.1088/1361-6382/ab3d4b}{\emph{Class. Quant. Grav.}
  {\bfseries 36} (2019) 195017},
  [\href{https://arxiv.org/abs/1905.00971}{{\ttfamily 1905.00971}}].

\bibitem{Alessio:2020ioh}
F.~Alessio, G.~Barnich, L.~Ciambelli, P.~Mao and R.~Ruzziconi, \emph{{Weyl
  charges in asymptotically locally AdS$_3$ spacetimes}},
  \href{http://dx.doi.org/10.1103/PhysRevD.103.046003}{\emph{Phys. Rev. D}
  {\bfseries 103} (2021) 046003},
  [\href{https://arxiv.org/abs/2010.15452}{{\ttfamily 2010.15452}}].

\bibitem{Fiorucci:2020xto}
A.~Fiorucci and R.~Ruzziconi, \emph{{Charge Algebra in Al(A)dS$_n$
  Spacetimes}},  [\href{https://arxiv.org/abs/2011.02002}{{\ttfamily
  2011.02002}}].

\bibitem{Adami:2020ugu}
H.~Adami, M.~M. Sheikh-Jabbari, V.~Taghiloo, H.~Yavartanoo and C.~Zwikel,
  \emph{{Symmetries at null boundaries: two and three dimensional gravity
  cases}}, \href{http://dx.doi.org/10.1007/JHEP10(2020)107}{\emph{JHEP}
  {\bfseries 10} (2020) 107},
  [\href{https://arxiv.org/abs/2007.12759}{{\ttfamily 2007.12759}}].

\bibitem{Ruzziconi:2020wrb}
R.~Ruzziconi and C.~Zwikel, \emph{{Conservation and Integrability in
  Lower-Dimensional Gravity}},
  [\href{https://arxiv.org/abs/2012.03961}{{\ttfamily 2012.03961}}].

\bibitem{Wieland:2020gno}
W.~Wieland, \emph{{Null infinity as an open Hamiltonian system}},
  [\href{https://arxiv.org/abs/2012.01889}{{\ttfamily 2012.01889}}].

\bibitem{Wieland:2021eth}
W.~Wieland, \emph{{Barnich-Troessaert Bracket as a Dirac Bracket on the
  Covariant Phase Space}},  [\href{https://arxiv.org/abs/2104.08377}{{\ttfamily
  2104.08377}}].

\bibitem{Freidel:2021yqe}
L.~Freidel, R.~Oliveri, D.~Pranzetti and S.~Speziale, \emph{{The Weyl BMS group
  and Einstein's equations}},
  \href{http://dx.doi.org/10.1007/JHEP07(2021)170}{\emph{JHEP} {\bfseries 07}
  (4, 2021) 170}, [\href{https://arxiv.org/abs/2104.05793}{{\ttfamily
  2104.05793}}].

\bibitem{anderson1992introduction}
I.~M. Anderson, \emph{Introduction to the variational bicomplex},
  {\emph{Contemporary Mathematics} {\bfseries 132} (1992) }.

\bibitem{Harlow:2019yfa}
D.~Harlow and J.-Q. Wu, \emph{{Covariant phase space with boundaries}},
  \href{http://dx.doi.org/10.1007/JHEP10(2020)146}{\emph{JHEP} {\bfseries 10}
  (2020) 146}, [\href{https://arxiv.org/abs/1906.08616}{{\ttfamily
  1906.08616}}].

\bibitem{Margalef-Bentabol:2020teu}
J.~Margalef-Bentabol and E.~J.~S. Villase\~nor, \emph{{Geometric formulation of
  the Covariant Phase Space methods with boundaries}},
  \href{http://dx.doi.org/10.1103/PhysRevD.103.025011}{\emph{Phys. Rev. D}
  {\bfseries 103} (2021) 025011},
  [\href{https://arxiv.org/abs/2008.01842}{{\ttfamily 2008.01842}}].

\bibitem{Compere:2018aar}
G.~Comp\`ere and A.~Fiorucci, \emph{{Advanced Lectures on General Relativity}},
   [\href{https://arxiv.org/abs/1801.07064}{{\ttfamily 1801.07064}}].

\bibitem{Papadimitriou:2005ii}
I.~Papadimitriou and K.~Skenderis, \emph{{Thermodynamics of asymptotically
  locally AdS spacetimes}},
  \href{http://dx.doi.org/10.1088/1126-6708/2005/08/004}{\emph{JHEP} {\bfseries
  08} (2005) 004}, [\href{https://arxiv.org/abs/hep-th/0505190}{{\ttfamily
  hep-th/0505190}}].

\bibitem{Compere:2008us}
G.~Compere and D.~Marolf, \emph{{Setting the boundary free in AdS/CFT}},
  \href{http://dx.doi.org/10.1088/0264-9381/25/19/195014}{\emph{Class. Quant.
  Grav.} {\bfseries 25} (2008) 195014},
  [\href{https://arxiv.org/abs/0805.1902}{{\ttfamily 0805.1902}}].

\bibitem{Freidel:2019ohg}
L.~Freidel, F.~Hopfm\"uller and A.~Riello, \emph{{Asymptotic Renormalization in
  Flat Space: Symplectic Potential and Charges of Electromagnetism}},
  \href{http://dx.doi.org/10.1007/JHEP10(2019)126}{\emph{JHEP} {\bfseries 10}
  (2019) 126}, [\href{https://arxiv.org/abs/1904.04384}{{\ttfamily
  1904.04384}}].

\bibitem{Speranza:2017gxd}
A.~J. Speranza, \emph{{Local phase space and edge modes for
  diffeomorphism-invariant theories}},
  \href{http://dx.doi.org/10.1007/JHEP02(2018)021}{\emph{JHEP} {\bfseries 02}
  (2018) 021}, [\href{https://arxiv.org/abs/1706.05061}{{\ttfamily
  1706.05061}}].

\bibitem{Kosmann}
Y.~Kosmann-Schwarzbach, \emph{The Noether Theorems}, pp.~55--64.
\newblock Springer New York, New York, NY, 2011.
\newblock 10.1007/978-0-387-87868-3.

\bibitem{kosmannschwarzbach2020noether}
Y.~Kosmann-Schwarzbach, \emph{The noether theorems in context},
  [\href{https://arxiv.org/abs/2004.09254}{{\ttfamily 2004.09254}}].

\bibitem{Noether:1918zz}
E.~Noether, \emph{{Invariant Variation Problems}},
  \href{http://dx.doi.org/10.1080/00411457108231446}{\emph{Gott. Nachr.}
  {\bfseries 1918} (1918) 235--257},
  [\href{https://arxiv.org/abs/physics/0503066}{{\ttfamily physics/0503066}}].

\bibitem{Bessel-H}
E.~Bessel-Hagen, \emph{Uber die erhaltungss\"atze der elektrodynamik},
  {\emph{Mathematische Annalen} {\bfseries 84} (1921) 258--276}.

\bibitem{Bessel-HT}
D.~H. Bessel-Hagen, E. Translated by~Delphenich, ``On the conservation laws of
  electrodynamics.''.

\bibitem{Tachikawa:2006sz}
Y.~Tachikawa, \emph{{Black hole entropy in the presence of Chern-Simons
  terms}}, \href{http://dx.doi.org/10.1088/0264-9381/24/3/014}{\emph{Class.
  Quant. Grav.} {\bfseries 24} (2007) 737--744},
  [\href{https://arxiv.org/abs/hep-th/0611141}{{\ttfamily hep-th/0611141}}].

\bibitem{Azeyanagi:2015gqa}
T.~Azeyanagi, R.~Loganayagam and G.~S. Ng, \emph{{Anomalies, Chern-Simons Terms
  and Black Hole Entropy}},
  \href{http://dx.doi.org/10.1007/JHEP09(2015)121}{\emph{JHEP} {\bfseries 09}
  (2015) 121}, [\href{https://arxiv.org/abs/1505.02816}{{\ttfamily
  1505.02816}}].

\bibitem{Brown:1986nw}
J.~D. Brown and M.~Henneaux, \emph{{Central Charges in the Canonical
  Realization of Asymptotic Symmetries: An Example from Three-Dimensional
  Gravity}}, \href{http://dx.doi.org/10.1007/BF01211590}{\emph{Commun. Math.
  Phys.} {\bfseries 104} (1986) 207--226}.

\bibitem{Troessaert:2015nia}
C.~Troessaert, \emph{{Hamiltonian surface charges using external sources}},
  \href{http://dx.doi.org/10.1063/1.4947177}{\emph{J. Math. Phys.} {\bfseries
  57} (2016) 053507}, [\href{https://arxiv.org/abs/1509.09094}{{\ttfamily
  1509.09094}}].

\bibitem{Iyer:1994ys}
V.~Iyer and R.~M. Wald, \emph{{Some properties of Noether charge and a proposal
  for dynamical black hole entropy}},
  \href{http://dx.doi.org/10.1103/PhysRevD.50.846}{\emph{Phys. Rev.} {\bfseries
  D50} (1994) 846--864}, [\href{https://arxiv.org/abs/gr-qc/9403028}{{\ttfamily
  gr-qc/9403028}}].

\bibitem{Gibbons:1976ue}
G.~W. Gibbons and S.~W. Hawking, \emph{{Action Integrals and Partition
  Functions in Quantum Gravity}},
  \href{http://dx.doi.org/10.1103/PhysRevD.15.2752}{\emph{Phys. Rev. D}
  {\bfseries 15} (1977) 2752--2756}.

\bibitem{Oliveri:2019gvm}
R.~Oliveri and S.~Speziale, \emph{{Boundary effects in General Relativity with
  tetrad variables}},
  \href{http://dx.doi.org/10.1007/s10714-020-02733-8}{\emph{Gen. Rel. Grav.}
  {\bfseries 52} (2020) 83},
  [\href{https://arxiv.org/abs/1912.01016}{{\ttfamily 1912.01016}}].

\bibitem{Bondi:1960jsa}
H.~Bondi, \emph{{Gravitational Waves in General Relativity}},
  \href{http://dx.doi.org/10.1038/186535a0}{\emph{Nature} {\bfseries 186}
  (1960) 535--535}.

\bibitem{BMS}
H.~Bondi, M.~G.~J. van~der Burg and A.~W.~K. Metzner, \emph{{Gravitational
  waves in general relativity. 7. Waves from axisymmetric isolated systems}},
  \href{http://dx.doi.org/10.1098/rspa.1962.0161}{\emph{Proc. Roy. Soc. Lond.}
  {\bfseries A269} (1962) 21--52}.

\bibitem{Sachs62}
R.~Sachs, \emph{{On the characteristic initial value problem in gravitational
  theory}}, {\emph{J.Math.Phys.} {\bfseries 3} (1962) 908--914}.

\bibitem{Campiglia:2014yka}
M.~Campiglia and A.~Laddha, \emph{{Asymptotic symmetries and subleading soft
  graviton theorem}},
  \href{http://dx.doi.org/10.1103/PhysRevD.90.124028}{\emph{Phys. Rev. D}
  {\bfseries 90} (2014) 124028},
  [\href{https://arxiv.org/abs/1408.2228}{{\ttfamily 1408.2228}}].

\bibitem{Ciambelli:2019lap}
L.~Ciambelli, R.~G. Leigh, C.~Marteau and P.~M. Petropoulos, \emph{{Carroll
  Structures, Null Geometry and Conformal Isometries}},
  \href{http://dx.doi.org/10.1103/PhysRevD.100.046010}{\emph{Phys. Rev. D}
  {\bfseries 100} (2019) 046010},
  [\href{https://arxiv.org/abs/1905.02221}{{\ttfamily 1905.02221}}].

\end{thebibliography}\endgroup

\end{document}